\documentclass[10pt, a4paper]{article}

\usepackage{amsmath}
\usepackage{amssymb}
\usepackage[T1]{fontenc}
\usepackage{graphicx}
\usepackage{authblk} %authors and affiliations
\usepackage[a4paper]{geometry}
\usepackage[font = small]{caption}
\usepackage{hyperref}

\title{Tuning the pseudospin polarization of graphene by a pseudo-magnetic field}
\date{}

\author[1]{Alexander Georgi}
\author[1]{Peter Nemes-Incze}
\author[2, 8]{Ramon Carrillo-Bastos}
\author[5, 9]{Daiara Faria}
\author[3, 6]{Silvia Viola Kusminskiy}
\author[8]{Dawei Zhai}
\author[3]{Martin Schneider}
\author[1]{Dinesh Subramaniam}
\author[4]{Torge Mashoff}
\author[1]{Nils M. Freitag}
\author[1]{Marcus Liebmann}
\author[1]{Marco Pratzer}
\author[7]{Ludger Wirtz}
\author[10]{Colin R. Woods}
\author[10]{Roman V. Gorbachev}
\author[10]{Yang Cao}
\author[10]{Kostya S. Novoselov}
\author[8]{Nancy Sandler}
\author[1, *]{Markus Morgenstern}

\affil[1]{II. Institute of Physics B and JARA-FIT, RWTH Aachen University, 52074 Aachen, Germany}
\affil[2]{Facultad de Ciencias, Universidad Aut\'{o}noma de Baja California, Apartado Postal 1880, 22800 Ensenada, Baja California, M\'{e}xico.} 
\affil[3]{Dahlem Center for Complex Quantum Systems and Institut f\"ur Theoretische Physik, Freie Universit\"at Berlin, 14195 Berlin, Germany}
\affil[4]{Institut f\"ur Physik, Johannes Gutenberg-Universit\"at, 55099 Mainz, Germany} 
\affil[5]{Instituto de F\'isica, Universidade Federal Fluminense, Niter\'oi, 24210-340 Rio de Janeiro, Brazil}
\affil[6]{Institute for Theoretical Physics II, University of Erlangen-N\"{u}remberg, 91058 Erlangen, Germany}
\affil[7]{Physics and Materials Science Research Unit, University of Luxembourg, L-1511 Luxembourg, Luxembourg}
\affil[8]{Department of Physics and Astronomy, Nanoscale and Quantum Phenomena Institute, Ohio University, Athens, 45701 Ohio, USA}
\affil[9]{Instituto Polit\'{e}cnico, Universidade do Estado do Rio de Janeiro, 28625-570 Nova Friburgo, Brazil.}
\affil[10]{School of Physics and Astronomy, University of Manchester, Manchester M13 9PL, United Kingdom}
\affil[*]{corresponding author: mmorgens@physik.rwth-aachen.de}

\begin{document}
\maketitle
%make line spacing from here on 2
\openup 0.5 em

\newpage

\textbf{
	One of the intriguing characteristics of honeycomb lattices is the appearance of a pseudo-magnetic field as a result of mechanical deformation.
	In the case of graphene, the Landau quantization resulting from this pseudo-magnetic field has been measured using scanning tunneling microscopy.
	Here we show that a signature of the  pseudo-magnetic field is a local sublattice symmetry breaking observable as a redistribution of the local density of states.
	This can be interpreted as a polarization of graphene's pseudospin due to a strain induced pseudo-magnetic field, in analogy to the alignment of a real spin in a magnetic field.
	We reveal this sublattice symmetry breaking by tunably straining graphene using the tip of a scanning tunneling microscope.
	The tip locally lifts the graphene membrane from a SiO$_2$ support, as visible by an increased slope of the $I(z)$ curves.
	The amount of lifting is consistent with molecular dynamics calculations, which reveal a deformed graphene area under the tip in the shape of a Gaussian.
	The pseudo-magnetic field induced by the deformation becomes visible as a sublattice symmetry breaking which scales with the lifting height of the strained deformation and therefore with the pseudo-magnetic field strength.
	Its magnitude is quantitatively reproduced by analytic and tight-binding models, revealing fields of 1000 T.
	These results might be the starting point for an effective THz valley filter, as a basic element of valleytronics.}\\

Strain engineering in graphene has been pursued intensely to modify its electronic properties \cite{Amorim2015, Pereira2009, Huang2011a, ZhangY2013},
mostly with a focus on deformations able to reproduce Landau level-like gapped spectra \cite{Guinea2009, Levy2010, Lu2012, Gomes2012}.
In addition to these effects, several theoretical works predict broken sublattice symmetry, measurable by the local density of states (LDOS) distribution in the presence of non-uniform strain \cite{Wehling2008a, Barraza-Lopez2013, Moldovan2013, Neek-Amal2013, Carrillo-Bastos2014, PachecoSanjuan2014, Schneider2015, Settnes2016}.
This local sublattice symmetry breaking (SSB) implies a valley filtering property in reciprocal space that may be exploited for valleytronic applications via a clever and controlled tuning of strain patterns \cite{Fujita2010, Low2010, Stegmann2015, Carrillo-Bastos2016, Settnes2016b, Milovanovic2016, Chaves2011}.

In the Dirac description, the sublattice degree of freedom is represented by a pseudospin, and a sublattice symmetry breaking is akin to a pseudospin polarization.
It is thus tempting to assign the strain related SSB to an alignment of the pseudospin that occurs in the presence of a pseudo-magnetic field \cite{Sasaki2008}.
Below we present an intuitive understanding of the phenomenon, using the squared Dirac Hamiltonian and explain it qualitatively and quantitatively by a coupling of the pseudospin to the pseudo-magnetic field that appears in the presence of strain.
The interpretation is corroborated by experiments that use the tip of a scanning tunneling microscope (STM) to deliberately strain a graphene sample locally, in the form of a small Gaussian bump, and at the same time to map the imbalance of the local density of states (LDOS) at the sublattice level.
Moreover, the measured sublattice contrast is quantitatively reproduced by an analytical model \cite{Schneider2015}.
These results provide a natural explanation for previous reports of SSB observed by STM in graphene, under non-tunable mechanical deformations \cite{Lu2012, Xu2009, Sun2009, Gomes2012}, which have so far not been attributed to a pseudospin polarization.

In the low-energy continuum Hamiltonian description for the electronic properties of graphene and other 2D-materials with a honeycomb lattice, mechanical deformations lead to a vector potential $\vec{{\cal{A}}}_{\mathrm{ps}}$, which is directly proportional to specific strain terms \cite{Kane1997, Suzuura2002, Vozmediano2010, Guinea2009, Cazalilla2014}.
The spatial dependence of $\vec{{\cal{A}}}_{\mathrm{ps}}$ critically influences the dynamics of charge carriers \cite{Pereira2009}.
A mechanical deformation with $\nabla \times \vec{{\cal{A}}}_{\mathrm{ps}} \neq 0$, results in an effective pseudo-magnetic field, perpendicular to the graphene plane $B_{\mathrm{ps}} = (\nabla \times \vec{{\cal{A}}}_{\mathrm{ps}})_z$ that couples with different sign to states in the two valleys \cite{Kane1997, Vozmediano2010}, i.e. it moves electrons in clockwise/counter-clockwise circles, respectively.
An effective way to analyze the effect of $B_{\mathrm{ps}}$ on the pseudospin degree of freedom is realized by squaring the Dirac Hamiltonian \cite{Aharonov1979, Sasaki2008}.
While the squared Hamiltonian describes the same physics as the original one, it also provides additional insight into the behavior of Dirac particles in a magnetic field.
Following this procedure, for both valleys we obtain (Supplement S2-1):
\begin{equation}
E^2 \Psi^{\mathrm{A}} = v_{\mathrm{F}}^2 \left[ \vec{\pi}^2 + e \hbar B_{\mathrm{ps}}\right]\Psi^{\mathrm{A}}
\end{equation}
\begin{equation}
E^2 \Psi^{\mathrm{B}} = v_{\mathrm{F}}^2 \left[ \vec{\pi}^2 - e \hbar B_{\mathrm{ps}}\right]\Psi^{\mathrm{B}}
\end{equation}
Here, $E$ is the energy, $\Psi^{\mathrm{A/B}}$ is the wave function amplitude on the corresponding sublattice $\mathrm{A/B}$, $\vec{\pi} = (\vec{p} \pm e \vec{{\cal{A}}}_{\mathrm{ps}})$ is the canonical momentum at each valley ($\pm$), $v_{\mathrm F}$ the Fermi velocity and $\vec{p}$ the momentum measured from the respective Dirac points $\mathrm{(K, K')}$.
The first term ($\vec{\pi}^{2}$) leads to Landau quantization, provided $B_{\mathrm{ps}}$ is homogeneous on the cyclotron radius scale, as observed by STM \cite{Levy2010, Lu2012, Gomes2012}.
The second term (with prefactor $v_{\mathrm F}^2 e \hbar = 658$ meV$^2$/T) corresponds to the coupling of $B_{\mathrm{ps}}$ to the graphene pseudospin.
It appears with opposite signs at sublattices A and B, shifting the energy of the respective states in opposite directions, thereby giving rise to a SSB, i.e. a pseudospin polarization.
The SSB is identical for both valleys since the change in sign of $B_{\mathrm{ps}}$ between valleys compensates the sign change in sublattice space (see Supplement S2).
An important feature of the $B_{\mathrm{ps}}$ - pseudospin coupling is its locality, that allows to use the SSB as a local fingerprint for even strongly inhomogeneous $B_{\mathrm{ps}}$ (strain).
The sublattice polarization resulting from the pseudo-magnetic field has been predicted in several theoretical works \cite{Wehling2008a, Barraza-Lopez2013, Moldovan2013, Neek-Amal2013, Carrillo-Bastos2014, PachecoSanjuan2014, Schneider2015, Settnes2016}.
The term that breaks the sublattice symmetry is sometimes referred to in the literature as pseudo-Zeeman coupling \cite{Sasaki2008,Katsnelson2012,Kim2011c}.
Its relation to the classic Zeeman effect for massive fermions becomes obvious after squaring the Dirac Hamiltonian and developing it for the non-relativistic limit \cite{Sakurai2010} (see Supplement S2-1).
Although for graphene the appropriate description is in terms of a massless Dirac equation, the analogy holds in the sense that the energy separation between the two pseudospin orientations is due to the coupling to the pseudo-magnetic field.
It is important to emphasize the difference between this term and another with the same expression, proposed as a gap opening perturbation for the Dirac Hamiltonian and unfortunately dubbed 'pseudo-Zeeman' term \cite{Manes2013}, since a Zeeman coupling breaks a degeneracy without necessarily opening a gap at the Dirac point.
As shown in Supplement section S2, in order to open a gap the pseudo-magnetic field should be even under inversion, while this is not the case for centrosymmetric deformations as the ones modeled in this work.

% our data ----------------------------------

To produce $B_{\mathrm{ps}}$ and measure the resulting SSB, we use the tip of a scanning tunneling microscope which is known to locally strain graphene due to attractive van der Waals (vdW) forces \cite{Mashoff2010, Klimov2012, Wolloch2015}.
Due to these forces, a Gaussian-shaped deformation forms below the W tip, locally lifting the graphene from its SiO$_2$ substrate (Fig. \ref{fig:intro}a), as evidenced by molecular dynamics calculations (see Supplement S3).
The deformation moves along with the tip while scanning (Fig. \ref{fig:intro}b, Supplementary Video).
It has typical dimensions of 5 \AA\ halfwidth and 1 \AA\ height.
The lifting height $H$ is tunable either by the tip-graphene distance $z$, adjusted by the tunneling current $I$, or by the locally varying adhesion forces of the substrate \cite{Mashoff2010}.
The mechanical strain within the Gaussian deformation results in a threefold symmetric $B_{\mathrm{ps}}$ pattern (color scale in Fig. \ref{fig:intro}c) \cite{Schneider2015}, which shifts the local density of states (LDOS) in opposite directions at each sublattice.
The resulting SSB, calculated by a nearest neighbor tight-binding model \cite{Carrillo-Bastos2014}, maps $B_{\mathrm{ps}}$ in terms of sign and strength down to the atomic scale (Fig. \ref{fig:intro}c) even while $B_{\mathrm{ps}}$ varies strongly on the scale of the pseudo-magnetic length (0.4 - 1 nm).
A consequence of this strong variation is the lack of Landau levels in tunneling spectroscopy curves.

The key to measure the SSB is to tunnel into areas of large $B_{\mathrm{ps}}$, i.e. a few atomic distances offset from the deformation centre.
The inherent asymmetry of real STM tips makes this the common situation.
We find SSB for $\sim$50\% of the individually prepared tips, hence the tunneling atom is adequately offset with respect to the force centre of the tip, i.e. the Gaussian maximum.
Here, we present results from a single tip showing the strongest SSB within our experiments.
However, comparable results are observed with other tips.
In particular, if the tip remains unchanged, the same sublattice appears brighter in all areas of the sample (Supplement, Fig. S9), matching with the expectation that one always probes the same local region of the Gaussian, i.e. the same sign of $B_{\mathrm{ps}}$ (Supplementary Video, Fig. \ref{fig:intro}b, c).
STM images in Fig. \ref{fig:intro}d-g demonstrate a controlled increase of SSB with increasing lifting force, i.e. increasing $I$, decreasing $z$, respectively.

Next, we show that the sublattice contrast $C = 2(\nu_{\mathrm{A}} - \nu_{\mathrm{B}})/(\nu_{\mathrm{A}} + \nu_{\mathrm{B}})$, with $\nu_{\mathrm{A/B}}$ the LDOS on sublattice A/B, can be related to $H$, the height of the Gaussian deformation \cite{Schneider2015}, due to the dependence $B_{\mathrm{ps}} \propto H^2$.
We estimate $H$ by comparing measured $I(Z)$ curves ($Z$: distance between tip apex and SiO$_2$), with the standard exponential decay expected from the tunneling model.
Therefore we use the work-functions of graphene $\Phi_{\mathrm G}$ = 4.6 eV \cite{Yu2009a} and tungsten $\Phi_{\mathrm W}$ = 5.3 eV \cite{Todd1973} (red line in Fig. \ref{fig:lift}a).
The measured curves follow the usual dependence at large $Z$, but strongly deviate at smaller $Z$.
Applying the tunneling model \cite{Tersoff1983}, the steepest areas (1-50 nA) would correspond to impossibly large work-functions of $\Phi$ = 140 eV (green) and $\Phi$ = 21 eV (blue).
This indicates the local lifting of graphene towards the tip, which increases $I$ beyond the expected increase due to the tip movement.
An elastic stretching of the tip is ruled out, since the same tips did not exhibit deviations from the tunneling model on Au(111).
Furthermore, since the slope of the $\mathrm{ln}(I(Z))$ curves changes during the approach and varies across the sample surface, the previously reported \cite{Zhang2008} slope change due to high momentum transfer during tunneling into the K points can also be ruled out.
The lifting amplitude $H_{\mathrm{exp}}$ is thus well estimated as the difference between the measured $I(Z)$ and the tunneling model (red line) as marked.
Variations of $H_{\mathrm{exp}}$ (green vs. blue curve) indicate variations in the adhesion to the substrate \cite{Geringer2009, Mashoff2010}.
Importantly, a map of the observed lifting heights $H_{\mathrm{exp}}$ (Fig. \ref{fig:lift}d) correlates with a map of the SSB (Fig. \ref{fig:lift}e).
SSB is consistently observed everywhere (Fig. S9 of the supplement) and is reversibly tunable on the same area (Fig. \ref{fig:lift}i).

In the following, we establish the relation between the apparent sublattice height difference $\Delta z$ and $H_{\mathrm{exp}}$.
We select areas of similar lifting height (for example blue area in Fig. \ref{fig:lift}d), subtract long-range corrugations and determine $\Delta z$ for each pair of neighboring atoms (Fig. \ref{fig:lift}e, Supplement S5).
Resulting histograms of $H_{\mathrm{exp}}$ and $\Delta z$ with indicated mean values $\langle H_{\mathrm{exp}} \rangle$ and $\langle \Delta z \rangle$ are shown for different $I$ in Fig. \ref{fig:lift}f, g.
The values of $\langle H_{\mathrm{exp}} \rangle$ and $\langle \Delta z \rangle$ recorded on different areas and at different $z$, i.e. $I$, collapse to a single curve (Fig. \ref{fig:lift}h).
Areas with larger $\langle H_{\mathrm{exp}} \rangle$ observed at the same $I$ are most likely caused by locally reduced adhesion to the SiO$_2$ \cite{Mashoff2010}, while the observed lower liftings of $\langle H_{\mathrm{exp}} \rangle \approx 1.5$ \AA\ are well reproduced by molecular dynamics calculations of graphene on flat, amorphous SiO$_2$ with an asymmetric W tip in tunneling distance (see Supplement S3).
Due to the much larger polarizability of W (21.4 \AA$^3$) with respect to Si (6.81 \AA$^3$) and O (0.7 \AA$^3$), the graphene is lifted from the SiO$_2$, even if the attractive dielectric forces between tip and graphene are neglected (Fig. \ref{fig:SSB}a-c).
Importantly, the graphene below the tip is well approximated by a Gaussian deformation.
The observed LDOS sublattice contrast can thus be compared with the predicted analytic expression \cite{Schneider2015}:
\begin{equation}
C_{\mathrm{theo}}(r, \theta) = -\frac{2 \beta H^2}{b a} \sin{(3 \theta)} g(r/b)
\label{eq:ctheo}
\end{equation}
($g(x) = \frac{1}{4x^3} [1 - e^{-2x^2} (1 + 2x^2 + 2x^4)]$, $\theta$: azimuthal angle, $r$: distance from centre, $\beta = 3$, $a$: lattice constant of graphene, $b$: width of the Gaussian deformation).
To compare with our experimental results, we determine the LDOS contrast $C_{\mathrm{exp}}$ from $\langle \Delta z \rangle$ (see Supplement S5) using:
\begin{equation}
C_{\mathrm{exp}} = 2 \frac{\mathrm{e}^{ \kappa \langle \Delta z \rangle} - 1}{\mathrm{e}^{\kappa \langle \Delta z \rangle} + 1}
\text{, with }
\kappa = \sqrt{\frac{8 m_{\mathrm{e}}}{\hbar^2} \left( \frac{\Phi_{\mathrm G} + \Phi_{\mathrm W}}{2} - \frac{e|V|}{2} \right)}
\label{eq:cexp},
\end{equation}
($m_{\mathrm{e}}$: free electron mass, $V$: the sample voltage).
The comparison is shown in Fig. \ref{fig:SSB}d with $C_{\mathrm{exp}}$ values being consistent with $C_{\mathrm{theo}}$ for deformation widths $b=5-7$ \AA, in excellent agreement with $b$ deduced from the molecular dynamics simulation (Fig. \ref{fig:SSB}a-c).
Finally, we examine the effect of the deformation being moved with the tip across the graphene lattice (see Supplementary video).
Figure \ref{fig:SSB}e, f displays the LDOS from a tight-binding (TB) calculation using two different central positions for the deformation, such that either sublattice A or B is imaged by the offset tunneling tip (black circle).
The lateral shift of the tip preserves the sign of the SSB, while the observed contrast changes slightly to 6.5\%\ (Fig. \ref{fig:SSB}g), from 6.9\%\ in the static deformation.
Conclusively, the model of pseudospin polarization describes our SSB data without any parameters which are not backed up by physical arguments.
Note that we have carefully considered tip artifacts and several alternative explanations for SSB, all of which strongly fail either quantitatively or qualitatively to explain the experimental data (Supplement S4).
Furthermore, strong SSB observed on a static graphene bubble (Supplement S7) further supports the straightforward pseudospin polarization scenario.

The observed pseudospin polarization dependent on $B_{\mathrm{ps}}$ adds an important ingredient to the analogy of graphene's Dirac charge carriers to ultrarelativistic particles.
In turn, the changes in SSB might be used to probe $B_{\mathrm{ps}}$ on small length scales \cite{Couto2014}.
Furthermore, the large values of $B_{\mathrm{ps}}$ ($\sim$1000 T) arising due to the dependence $B_{\mathrm{ps}}\propto H^2/b^3$ \cite{Schneider2015}, suggest the use of the strained region as a valley filter \cite{Settnes2016b, Milovanovic2016}, operating on nanometer length scales and switchable with THz frequency (Supplement S8).
Recently, valley currents with relaxation length of up to 1 $\mu$m have been measured \cite{Gorbachev2014, Shimazaki2015}, but so far only in static configurations.
Finally, the existence of strain induced SSB provides the first direct experimental evidence of the unique time reversal invariant nature of $B_{\mathrm{ps}}$.
Being fundamentally different from a real magnetic field, this property could provide novel ground states dominated by many body interactions not achievable otherwise \cite{Roy2014a, Ghaemi2012}, or in combination with a real magnetic field of comparable magnitude could mimic the decoupling of a chiral flavour as observed in the weak interaction \cite{Sasaki2010}.

\section*{Acknowledgement}
	We acknowledge discussions with M. I. Katsnelson, A. Bernevig, M. Kr\"{a}mer, W. Bernreuther, F. Libisch, C. Stampfer and C. Wiebusch,  assistance at the STM measurements and sample preparation by C. Pauly, C. Saunus, S. Hattendorf, V. Geringer. We acknowledge financial support by the Graphene Flagship (Contract No. NECT-ICT-604391) and the German Research Foundation via Li 1050/2-2 (A.G., P.N.I., M.P., M.L. and M.M.); DFG SPP 1459 and the A. v H. Foundation (M.S., S.V.K.); CNPq No.150222/2014-9 (D.F.); NSF No. DMR-1108285 (D.F., R.C-B., D.Z. and N.S.); PRODEP 2016 (R.C.-B).

{\small
\section*{Author contributions}
AG carried out the STM measurements assisted by DS. Molecular dynamics simulations were conducted by PNI and AG. Tight-binding calculations were done by RC-B and DF under the supervision of NS. Continuum model calculations were performed by DF, DZ, SVK, and MS with the help of NS. LW provided the DFT calculations for highly strained graphene. CRW, YC and RG prepared the graphene on BN sample, NMF measured it in STM. PNI, AG, MM and NS prepared the manuscript. All authors contributed to the data analysis and the revising of the manuscript. MM provided the general idea of the experiment and coordinated the project together with AG.
}

%---------------------------------------------------------------

\newpage

%Figures -------------------------------------------------------

\begin{figure}[h!]
\includegraphics[width = 1 \textwidth]{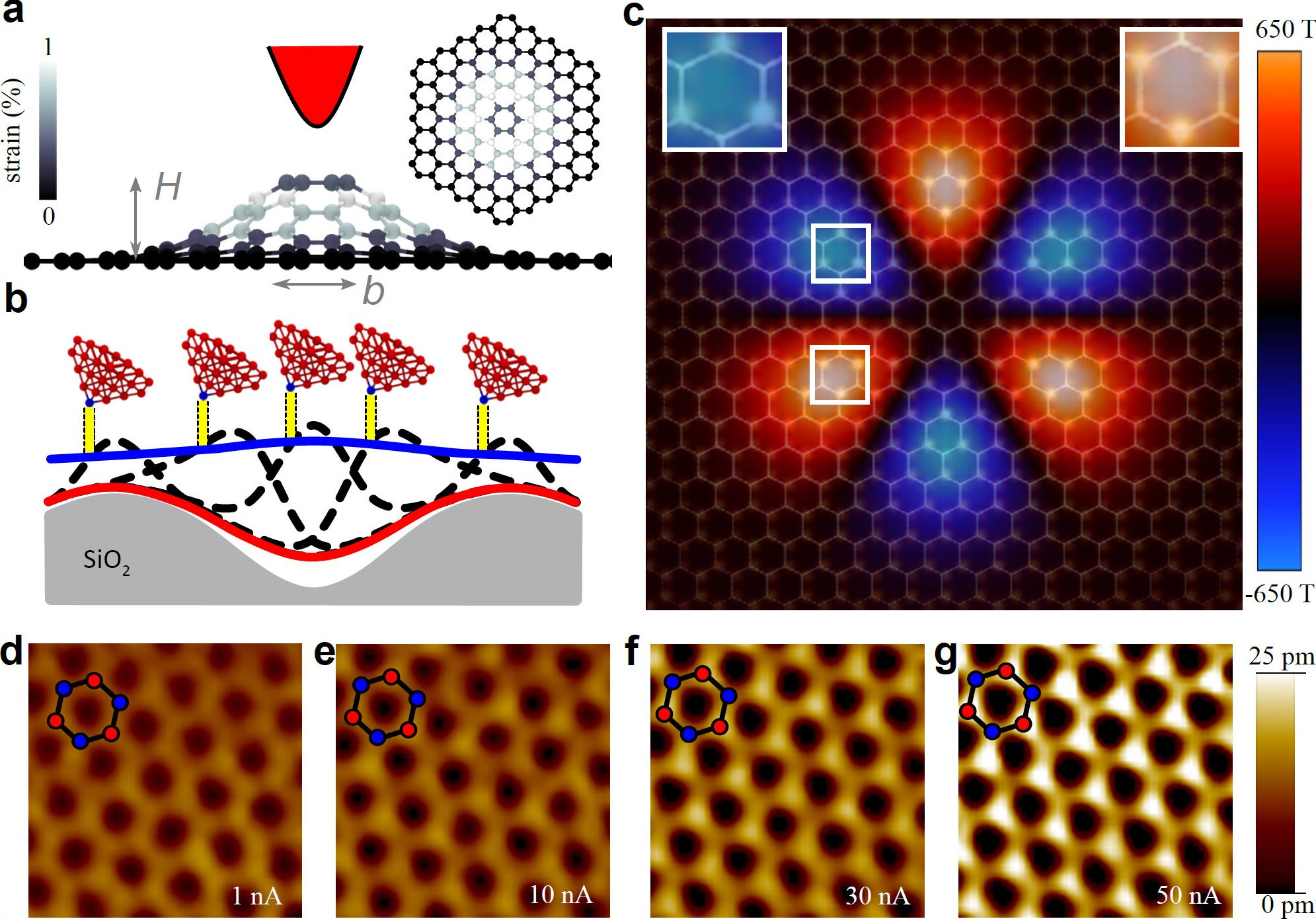}
\caption{\textbf{Sublattice symmetry breaking (SSB):}
\textbf{(a)}, Gaussian deformation induced in graphene by the attractive van der Waals force of the STM tip (amplitude $H$ = 1 \AA, width $b$ = 5 \AA) as observed in molecular dynamics calculation of graphene on SiO$_2$ (Fig. \ref{fig:SSB}a-c). Colour code represents the induced strain.
\textbf{(b)}, The deformation (black dashed line) follows the scanning STM tip (red balls), leading to the apparent STM image (blue) lifted with respect to the relaxed one (red line). Yellow bar represents the tunnelling current.
\textbf{(c)}, Colour code: pseudo-magnetic field pattern of the Gaussian deformation of (a) \cite{Schneider2015}. Honeycomb lattice is overlayed with LDOS magnitude of individual atoms as calculated in nearest neighbour tight-binding \cite{Carrillo-Bastos2014} marked by the brightness of the corresponding dots. White squares show areas of maximum $B_{\mathrm{ps}}$ magnified as insets.
\textbf{(d-g)}, Constant current STM images of the same graphene area on a SiO$_2$ substrate recorded at varying currents as marked ($T$ = 6 K, 1.3$\times$1.3 nm$^2$, $V$ = 0.5 V). A sketch of the graphene honeycomb lattice is overlaid with the different sublattices indicated by blue and red dots.
}
\label{fig:intro}
\end{figure}

\newpage

\begin{figure}[h!]
\includegraphics[width = 1 \textwidth]{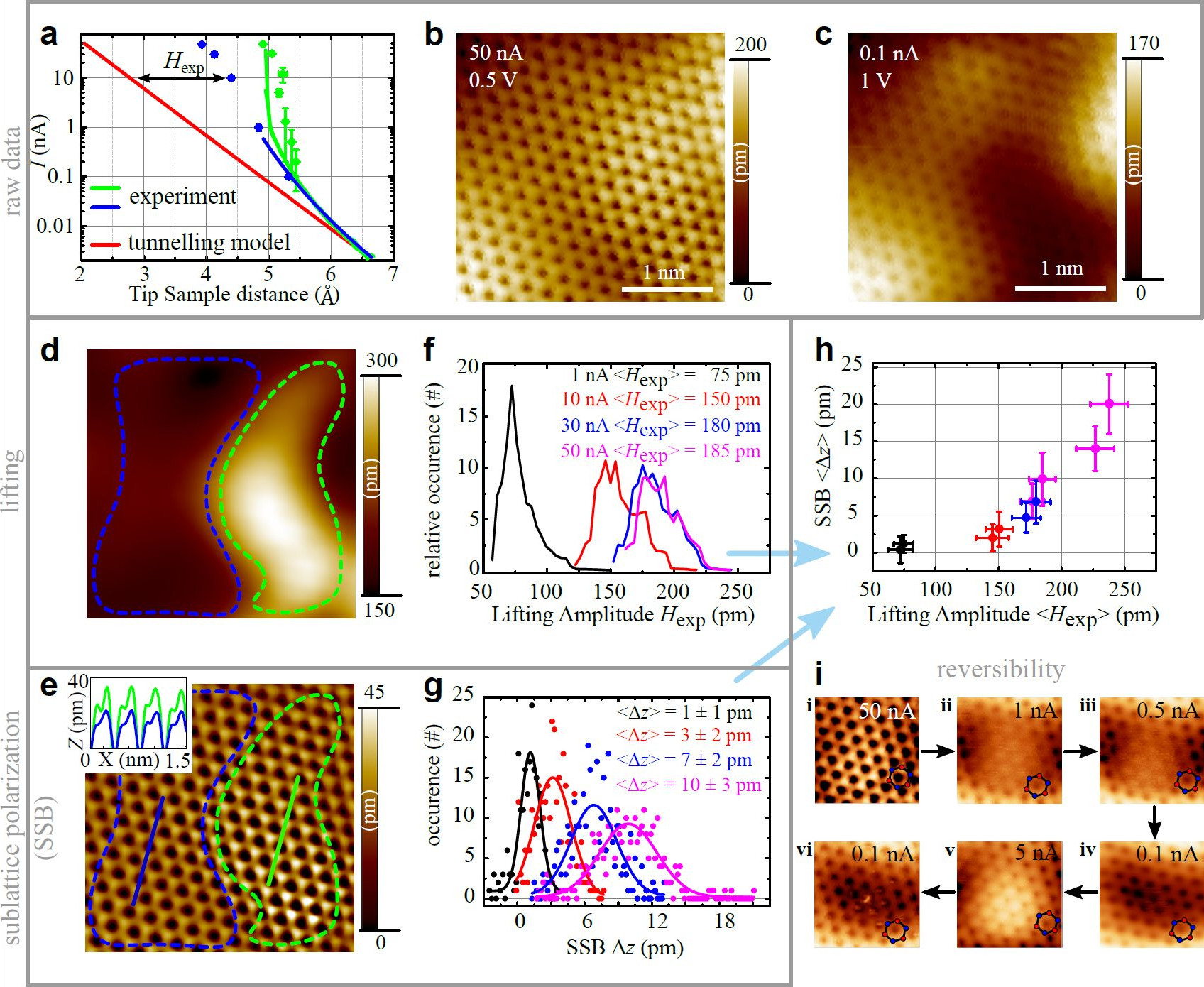}
\caption{\textbf{Relating lifting and SSB:}
\textbf{(a)}, $I(Z)$-curves with logarithmic $I$ scale for regions with low/high lifting (see (d) blue/green areas). Dots (lines) are measured with (without) STM feedback loop (Supplement S4-2). Red line results from a 1D vacuum tunnelling model \cite{Tersoff1983} using the work functions of graphene and tungsten. Black arrow indicates the deduced lifting amplitude $H_{\mathrm{exp}}$ of graphene.
\textbf{(b, c)}, Raw STM data of the same area at different $I$, $V$, as marked.
\textbf{(d)}, Lifting amplitude $H_{\mathrm{exp}}$ at $I$ = 50 nA, $V$ = 0.5 V deduced from (b, c) (Supplement S4).
\textbf{(e)}, Atomic corrugation of (b) obtained by subtracting the long-range morphology. Full lines mark the profile lines shown in the inset. Same curved, dashed lines in (d) and (e) (Supplement S4).
\textbf{(f)} Histogram of $H_{\mathrm{exp}}$ for the blue area in (d), for different $I$ as marked, $V=0.5$ V.
\textbf{(g)}, Histogram of $\Delta z$ for the same area, at different currents $I$ (points) with Gaussian fits (lines).
\textbf{(h)}, Measured $\langle \Delta z \rangle$ with respect to $\langle H_{\mathrm{exp}} \rangle$. Colours correspond to the accordingly coloured tunnelling current in (f). Data from other sample areas than (b, c) are included.
\textbf{(i)}, STM images of the same area (V = 0.5 V, 1.5$\times$1.5 nm$^2$) recorded consecutively as marked by the arrows at varying $I$.
}
\label{fig:lift}
\end{figure}

\newpage

\begin{figure}[h!]
\includegraphics[width = 1 \textwidth]{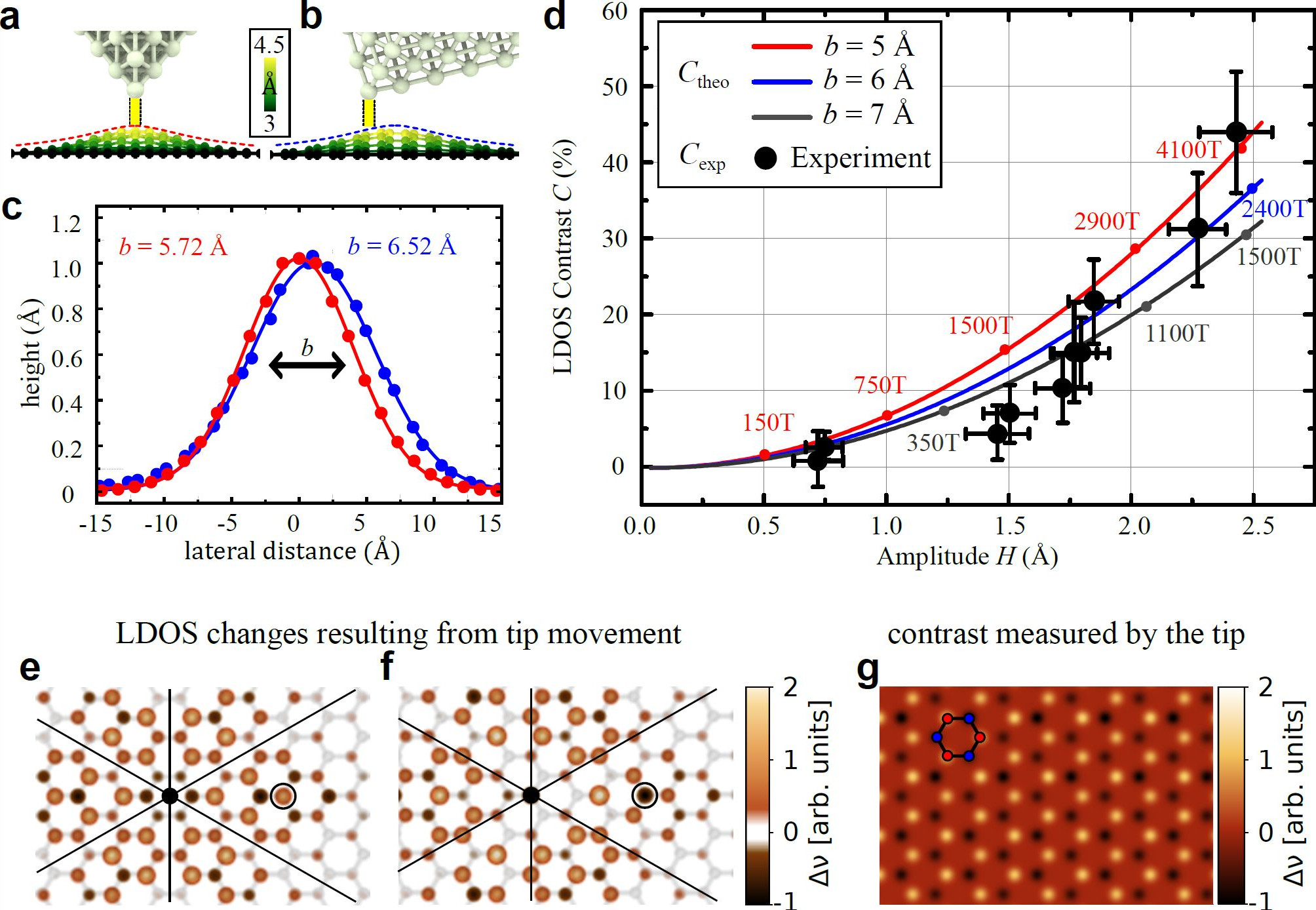}
\caption{\textbf{Comparing measured and calculated SSB:}
\textbf{(a, b)} Atomic configuration from molecular dynamics simulations of graphene on a SiO$_2$ substrate in presence of a pyramidal W (110) tip tilted by $\alpha$ = 30$^o$ with respect to the substrate normal, viewed from two perpendicular directions. The tunnelling current is visualized as a yellow bar, green colour code marks graphene distance relative to the substrate, smallest tip sample distance: $z$ = 3 \AA.
\textbf{(c)}, Atomic positions, plotted as dots, along the dashed lines in (a), (b) with Gaussian fits $h(r) = H \cdot \exp{(-r^2/b^2)}$ (lines).
\textbf{(d)}, Calculated and measured LDOS contrast $C$ as a function of the lifting amplitude $H$. $C_{\mathrm{theo}}$ results from the continuum model (Eq. (\ref{eq:ctheo})) and is plotted at $r = b$. $C_{\mathrm{exp}}$ results from the data of Fig. \ref{fig:lift}h using Eq. (\ref{eq:cexp}). The corresponding maximum pseudo-magnetic fields $B_{\mathrm{ps, max}}$ are marked at the dots on the full lines, in according colour. All data are recorded with the same tip showing maximum SSB within our experiments.
\textbf{(e, f)}, LDOS patterns ($\Delta \nu$ = $\nu_{\mathrm{deformation}}$ - $\nu_{\mathrm{flat}}$) resulting from TB calculation for a Gaussian deformation with $H$ = 1 \AA, $b$ = 5 \AA, and for different positions of the Gaussian centre (black dot) with respect to the graphene lattice. Black ring corresponds to the tunnelling position of the tilted tip being on sublattice A (e) and B (f).
\textbf{(g)}, Scanned LDOS pattern as observed by the tunnelling atom at the black ring in (e), (f).
}
\label{fig:SSB}
\end{figure}

\newpage

% ------------------------ Supplement -----------------------

	\title{\LARGE{\textbf{Supplementary information for:\\
	Tuning the pseudospin polarization of graphene by a pseudo-magnetic field}}}
	\date{}

	\maketitle

	% \newpage
	\tableofcontents
	%\listoftodos[Notes]

	% \openup 0.5em

\newcounter{supplement}
\newcounter{alter}[supplement]
\renewcommand{\thefigure}{S\arabic{figure}}
\renewcommand{\thepage}{S\arabic{page}}

\newpage

\refstepcounter{supplement}
\section*{S\arabic{supplement}: Experimental methods}
\label{sec:expmeth}
\addcontentsline{toc}{section}{S\arabic{supplement}: Experimental methods}

Graphene was prepared by mechanical exfoliation on SiO$_2$.
The layer thickness is determined by Raman spectroscopy.
To avoid surface contamination the graphene flake is contacted by microsoldering with indium leads \cite{Girit2007}.
For further cleaning, the sample was annealed at 100$^\circ$C in ultra-high vacuum (UHV) before it was cooled down to 5 K within our home-build UHV-STM system \cite{Mashoff2009}.
For STM measurements, an electrochemically etched tungsten tip is aligned to the sample by a long range optical microscope, using the indium contacts as cross hairs.

\refstepcounter{supplement}
\section*{S\arabic{supplement}: Relation between chiral symmetry, parity and sublattice symmetry breaking in graphene}
\label{sec:Diractheory}
\addcontentsline{toc}{section}{S\arabic{supplement}: Relation between chiral symmetry, parity and sublattice symmetry breaking in graphene}

Here we analyse the general requirements for a strain induced gauge field $\vec{{\cal{A}}}$ to produce a sublattice symmetry breaking (SSB). We show that $\vec{{\cal{A}}}$ will always produce a SSB as long its curl is non zero ($\nabla \times \vec{{\cal{A}}} \neq 0$), irrespective of the spatial mirror symmetry properties of $\vec{{\cal{A}}}$.

We start setting up the problem by considering the tight-binding model of graphene and the possible symmetry operations therein. The standard description of electron dynamics in pristine graphene is given by the tight-binding Hamiltonian:
\begin{equation}
H_{0} = -t \sum_{\vec{n}, \vec{\delta}_{i}} \left[ a^{\dagger}_{\vec{n}} \;b_{\vec{n}+\vec{\delta}_{i}} + c.c.\right]
\label{eq:TB}
\end{equation}
where $t$ is the nearest neighbour hopping parameter, $a_{\vec{n}}$ and $b_{\vec{n} +\vec{\delta}}$ are electron operators for atoms in sublattices $\mathrm{A}$ and $\mathrm{B}$ respectively, and $\vec{\delta}_{i}$ indicates the three lattice vectors that connect an atom in sublattice $\mathrm{A}$ with its nearest neighbour atoms in sublattice $\mathrm{B}$.
In pristine graphene the magnitude of $\vec{\delta}_{i}$ is the inter-atomic bond length $a$, and the three directions are related to each other by $120^{\circ}$ rotations around an axis perpendicular to the graphene plane.
This model fulfils the symmetries of the honeycomb lattice that includes these rotations and mirror reflections about three planes parallel to the three different carbon-carbon bond directions. Notice that all these symmetry operations relate sites on the same sublattice.
Because both sublattices are populated by the same type of atom, there are additional symmetries that hold when an exchange of sublattice sites is included.
This symmetry appears  in the literature as 'inversion' (or 'parity' in quantum field theory), and consists of two operations: an inversion of real space coordinates $\vec{r} \rightarrow -\vec{r}$ and a sublattice exchange $\mathrm{A} \rightarrow \mathrm{B}$ \cite{Gusynin2007, Zuelicke2012}.
For simplicity the centre of inversion can be thought of either the middle of the carbon-carbon bond or the centre of a hexagonal unit cell.
To analyse the consequences of this inversion symmetry in the presence of deformations, we use the spinor representation for wave functions at sites $\mathrm{A}$ and $\mathrm{B}$ in momentum space.
Introducing the Fourier transform of the operators $(a_{\vec{n}}; b_{\vec{n}+\vec{\delta}})$ defined in Eq.~\ref{eq:TB}, the Hamiltonian in this basis takes the form:
\begin{eqnarray}
H_{0} &=& \int_{BZ}\frac{d^{2}q}{(2\pi)^{2}} \Psi^{\dagger}(\vec{q}){\cal{H}}_{0}\Psi(\vec{q}) \\ \nonumber
{\cal{H}}_{0} &=&  
\left( 
\begin{array}{lr}
0  & \phi(\vec{q}) \\
\phi^{*}(\vec{q})& 0 \\
\end{array}
\right); \;\;\;\;\;   
\Psi(\vec{q})=
\left( 
\begin{array}{l}
\psi_{\mathrm{A}}(\vec{q}) \\
\psi_{\mathrm{B}}(\vec{q}) \\
\end{array}
\right) 
\end{eqnarray}
where the integral runs over the Brillouin zone, $\phi(\vec{q}) = -t \sum_{\vec{\delta}_{i}}e^{i\vec{q}\vec{\delta}_{i}}$ and $\vec{q}$ is measured with respect to the $\Gamma$ point.
The low-energy physics is obtained by expanding $\phi(\vec{q})$ around the two inequivalent reciprocal lattice points K and K' (valleys) and results in an effective Dirac Hamiltonian.
Using the basis $\Psi = (\psi^{\rm K}_{\rm A}, \psi^{\rm K}_{\rm B}, \psi^{\rm K'}_{\rm B}, \psi^{\rm K'}_{\rm A})$ defined by the corresponding Fourier components $\left( \psi_{\mathrm{A}}(\vec{p}); \psi_{\mathrm{B}}(\vec{p}) \right)$ expanded around these points \cite{Jackiw2007}:
\begin{equation}
\begin{array}{lcl}
\psi^{\rm K,K'}_{\rm A}(\vec{r}) = \int d^2 p e^{-i \vec{p} \vec{r}} a_{\rm K,K'}(\vec{p})
\\
\psi^{\rm K,K'}_{\rm B}(\vec{r}) = \int d^2 p e^{-i \vec{p} \vec{r}} b_{\rm K,K'}(\vec{p})
\end{array}
\label{eq:chir3}
\end{equation}
the Hamiltonian can be written as
\begin{equation}
\displaystyle{\cal{H}}_{0}
= v_{\rm F}
\begin{pmatrix}
\vec{\sigma} \cdot \vec{p} & 0 \\
0 & -\vec{\sigma} \cdot \vec{p}
\end{pmatrix}
\label{eq:H0}
\end{equation}
where $\vec{\sigma}$ is the vector of Pauli matrices acting on the sublattice spinor space, the momentum $\vec{p}$ is measured with respect to $\mathrm{K}$ and $\mathrm{K'}$ respectively and $\displaystyle{\cal{H}}_{0}$ acts on the 'valley' spinor space.
These expressions are derived from a real space coordinate frame $(x,y)$ such that $x$ is along the zigzag direction (perpendicular to the carbon-carbon bond).

By an appropriate transformation, Eq.~\ref{eq:H0} can be written in the chiral (Weyl) representation where the Hamiltonian is diagonal.
The eigenstates can be classified by energy $E$, momentum $\vec{p}$ and the quantum number $(\pm1)$ associated with a 'pseudo-helicity' operator defined as the identity in valley space and as $\Sigma_{\mathrm{ps}} = \vec{\sigma}\cdot\vec{p}/|p|$ in sublattice space.
We refer to it as 'pseudo-helicity' to differentiate it from the helicity operator defined in quantum field theory that refers to rotations in spin space \cite{Gusynin2007}.
The eigenstates or chiral states given in the $\Psi$ basis are:
\begin{eqnarray}
|1\rangle = |+E,  \vec{p}, +1\rangle & = & (e^{-i\theta(\vec{p})/2}, e^{i\theta(\vec{p})/2}, 0, 0)^{T}; \nonumber \\
|2\rangle = |-E, \vec{p}, -1\rangle & = & (e^{-i\theta(\vec{p})/2}, -e^{i\theta(\vec{p})/2}, 0, 0)^{T}; \nonumber \\
|3\rangle = |-E, \vec{p}, +1\rangle & = & (0, 0, e^{-i\theta(\vec{p})/2}, e^{i\theta(\vec{p})/2})^{T}; \nonumber \\
|4\rangle = |+E, \vec{p}, -1\rangle & = & (0, 0, e^{-i\theta(\vec{p})/2}, -e^{i\theta(\vec{p})/2})^{T}.
\label{eq:chiralstates}
\end{eqnarray}
Here $E$ refers to the energy and $\theta(\vec{p}) = \tan^{-1}(p_{y}/p_{x})$. In the chiral basis, wave function amplitudes are equal at both sublattices, and the system is said to exhibit chiral symmetry.

To analyse the role of deformations on the symmetries of the Hamiltonian it is convenient to use a covariant notation. We introduce the $\gamma$ matrices as: $ \gamma^i = \beta \alpha^i $, $ \gamma^0 = \beta $, $ \gamma^5 = i \gamma^0 \gamma^1 \gamma^2 \gamma^3 $ ($ i = 1,2,3$), with
\begin{equation}
\alpha^i =
\begin{pmatrix}
\sigma^i & 0 \\
0 & -\sigma^i
\end{pmatrix},\;\;\;
\beta =
\begin{pmatrix}
0 & {\cal{I}} \\
{\cal{I}} & 0
\end{pmatrix}
\label{eq:chir3.1}
\end{equation}
and ${\cal{I}}$ the $2 \times 2$ unit matrix. With these definitions, the graphene Hamiltonian density reads:
\begin{equation}
{\cal{H}}^{\mathrm{D}}_{0} = \bar{\Psi}(\vec{r}) \gamma^{\mu} p_{\mu} \Psi(\vec{r})
\label{eq:H0Dirac}
\end{equation}
with an implicit sum over $\mu =(1,2)$, $\bar{\Psi} = \Psi^{\dag} \gamma^0$is the Dirac adjoint spinor, and $p_{\mu} = -i\hbar \partial_{\mu}$.

The representation of the inversion operation is given in this notation by ${\cal{P}} = i \gamma^{0}P$, where $P$ executes the transformation $\vec{r} \rightarrow -\vec{r}$ \cite{Gusynin2007}.
A straightforward calculation shows that ${\cal{H}}^{D}_{0}$ is invariant under inversion.
For a given energy, its action on the chiral basis results in the exchange of states  with wave function amplitudes at different valleys (for example, it exchanges $|1(\vec{p})\rangle$ with $|4(\vec{p})\rangle$). 

A deformation in graphene affects the lattice vectors $\vec{\delta}_{i}$ and introduces a change in the hopping matrix elements $t' = t +\Delta t$ in Eq.~\ref{eq:TB}.
The terms including $\Delta t$ result in an effective gauge field $\vec{{\cal{A}}}$ \cite{Vozmediano2010, Sasaki2008, Barraza-Lopez2013}.
In the continuum model, and for small deformations these changes are described within elasticity theory by introducing the strain tensor $\varepsilon_{ij}= 1/2(\partial_{i}u_{j} + \partial_{j}u_{i}+\partial_{i}h \partial_{j}h)$ where $(u_{i}, h)$ are the in-plane and out-of-plane atomic displacements respectively.
The components of the effective gauge field then read $({\cal{A}}_{x}, {\cal{A}}_{y}) = \frac{\hbar \beta}{2{\it{a}}}(\varepsilon_{xx} - \varepsilon_{yy}, -2\varepsilon_{xy})$ with $\beta \sim 3$. Inclusion of such a term in Eq.~\ref{eq:H0} produces:
\begin{equation}
H
= v_{\rm F}
\begin{pmatrix}
\vec{\sigma}\cdot (\vec{p} - \vec{{\cal{A}}}) & 0 \\
0 & -\vec{\sigma} \cdot(\vec{p} + \vec{{\cal{A}}})
\end{pmatrix}
\label{eq:HA}
\end{equation}
When written in the covariant notation, the new Hamiltonian density reads:
\begin{equation}
{\cal{H}}^{\mathrm{D}} = \bar{\Psi} (\vec{r}) \gamma^{\mu} (p_{\mu} -\gamma^{5} {\cal{A}}_{\mu}) \Psi(\vec{r})
,\;\;\;\;\;
\gamma^{5} = 
\begin{pmatrix}
\cal{I} & 0 \\
0 & -\cal{I}
\end{pmatrix}
\label{eq:HDirac}
\end{equation}
Notice that the presence of the matrix $\gamma^{5}$ ensures the correct change of sign in the components of $\vec{{\cal{A}}}(\vec{r})$ by changing from K to K', as inherited from the lattice expressions obtained for $\Delta t$.

Now, let us consider how the specific spatial dependence of $\vec{{\cal{A}}}(\vec{r})$ influences the invariance of the Hamiltonian under inversion.
For the total Hamiltonian density to remain invariant under inversion, it must be an even function of $\vec{r}$, i.e. $P\vec{{\cal{A}}}(\vec{r})P^{-1} =\vec{{\cal{A'}}}(\vec{r'}) = \vec{{\cal{A'}}}(-\vec{r}) = \vec{{\cal{A'}}}(\vec{r})$ (axial vector) \cite{Jackiw2007}, as in the case of a perfect Gaussian deformation.
A reflection of this can be seen in Fig. 1c of the main text, where the graphene LDOS is symmetric with respect to inversion around the deformation centre.
Additionally, a deformation with odd gauge field, $\vec{{\cal{A'}}}(-\vec{r}) = -\vec{{\cal{A'}}}(\vec{r})$ (polar vector) breaks parity.
Invariance under inversion and time-reversal symmetries protects the degeneracy at the Dirac points.
If $\vec{{\cal{A}}}(\vec{r})$ is an odd function under inversion of space-coordinates, parity is broken and a gap opens at the Dirac points in addition to the chiral symmetry breaking.

Lastly we ask the question: what are the requirements for $\vec{{\cal{A}}}(\vec{r})$ to produce a SSB?
The gauge field term in $\cal{H}^{\mathrm{D}}$ represents an interaction added to ${\cal{H}}_{0}^{\mathrm{D}}$ that may or may not commute with it.
The commutator involves terms like $[\gamma^{\mu}p_{\mu}, \gamma^{\nu}\gamma^{5}{\cal{A}}_{\nu}]$ that are proportional to $\nabla \times \vec{{\cal{A}}}$.
If $\nabla \times \vec{{\cal{A}}} = 0$, ${\cal{H}}_{0}^{\mathrm{D}}$ and $\cal{H}^{\mathrm{D}}$ commute, and the gauge field $\vec{{\cal{A}}}$ can be removed from Eq.~\ref{eq:HDirac} by an appropriate gauge transformation. As a consequence, chiral symmetry is preserved and electronic densities will exhibit sublattice symmetry when imaged.
However, if $\nabla \times \vec{{\cal{A}}} \neq 0$, the commutator does not vanish and an effective 'pseudo-magnetic field' $B_{\mathrm{ps}}(\vec{r}) = (\nabla \times \vec{{\cal{A}}})_z$ is produced.
This pseudofield couples to the sublattice spinor producing an effective pseudospin polarization that selects the same sublattice at each valley as a straightforward calculation shows (see S\ref*{sec:Diractheory}-\ref{sec:zeeman}).

In conclusion, a deformation that induces a pseudo gauge field with $\nabla \times \vec{{\cal{A}}} \neq 0$ will induce a sublattice symmetry breaking irrespective of its specific functional dependence under inversion.

\refstepcounter{alter}
\subsection*{S\arabic{supplement}-\arabic{alter}: Link between pseudospin polarization and the Zeeman effect for massive particles}
\addcontentsline{toc}{subsection}{S\arabic{supplement}-\arabic{alter}: Link between pseudospin polarization and the Zeeman effect for massive particles}
\label{sec:zeeman}

To establish the connection between sublattice symmetry breaking and a pseudo-Zeeman coupling, we study the non-relativistic limit of the squared Dirac Hamiltonian \cite{Sasaki2008, Sakurai2010}. Using the $\Psi$ basis defined above and Eq.~\ref{eq:HA}, it reads:

\begin{equation}
H^2 = v_{\rm F}^2
\begin{pmatrix}
\vec{\sigma}(\vec{p}-\vec{{\cal{A}}})\cdot  \vec{\sigma}(\vec{p}-\vec{{\cal{A}}})& 0  \\
0 & [-\vec{\sigma}(\vec{p}+\vec{{\cal{A}}})][-\vec{\sigma}(\vec{p}+\vec{{\cal{A}}})] \\
\end{pmatrix}
\label{eq.:H2}
\end{equation}

Using the standard identity $(\vec{\sigma}\vec{X})\cdot(\vec{\sigma}\vec{Y}) = \vec{X}\cdot\vec{Y} + i \vec{\sigma}(\vec{X}\times\vec{Y})$, we obtain:
\begin{equation}
H^2 = v_{\rm F}^2
\begin{pmatrix}
(\vec{p} -\vec{{\cal{A}}})^{2} {\cal{I}} - B_{\mathrm{ps}}\sigma_{z}& 0  \\
0 & (\vec{p} + \vec{{\cal{A}}})^{2}{\cal{I}} + B_{\mathrm{ps}}\sigma_{z} \\
\end{pmatrix}
\label{eq.:H2LL}
\end{equation}
where we used $\vec{p} \times \vec{{\cal{A}}}(\vec{r}) = -i \nabla \times\vec{{\cal{A}}}(\vec{r}) -\vec{{\cal{A}}}(\vec{r}) \times \vec{p}$, with $\nabla \times\vec{{\cal{A}}}(\vec{r}) = e \hbar B_{\rm ps}\sigma_{z}$. The terms $v_{\rm F}^2 (\vec{p} \pm \vec{{\cal{A}}})^{2} {\cal{I}}$ correspond to the kinetic or orbital energy leading to pseudo-Landau levels in case of homogeneous $B_{\rm ps}$, while $v_{\rm F}^2  B_{\rm ps}\sigma_{z}$ is equivalent to a pseudo-Zeeman coupling term \cite{Sasaki2008,Katsnelson2012,Kim2011c}.
Its prefactor is $v_{\rm F}^2 e \hbar\simeq 658$ meV$^2$/T, i.e., the pseudo-Zeeman energy scales with the square root of the field.

Notice that the effect of the Pauli matrix $\sigma_{z}$ is to change the sign of $B_{\rm ps}$ at each sublattice within the same valley \cite{Kane1997,Suzuura2002}.
Thus, the sign change of $B_{\rm ps}$ between the two valleys compensates the change of sign between sublattices in each valley:
\begin{equation}
v_{\rm F}^2 \left( \pi^{2}_{K,K'} - e \hbar B_{\rm ps} \right) \psi_{\rm A}^{\rm K, K'} = E^2 \psi_{\rm A}^{\rm K, K'}
\label{eq:4}
\end{equation}
\begin{equation}
v_{\rm F}^2 \left( \pi^{2}_{K,K'}  + e \hbar  B_{\rm ps} \right) \psi_{\rm B}^{\rm K, K'} = E^2 \psi_{\rm B}^{\rm K, K'}
\label{eq:5}
\end{equation}
Here we used $\pi^{2}_{K,K'}$  to represent the orbital kinetic energy operator at each valley. 
For an arbitrary shape of the pseudo-magnetic field, the pseudo-Zeeman term locally shifts the LDOS upwards in energy on sublattice B and downwards in energy on sublattice A.
Thus, it leads to a sublattice polarization, akin to the spin polarization induced by a real magnetic field.
From the eigenvalue expression above we can also see that the pseudospin polarization is symmetric in energy, i.e. the LDOS increases for the same sublattice for both electrons and holes.

The magnetic field - spin interaction term $e \hbar v_{\rm F}^2 B_{\rm ps}$ for relativistic spin 1/2 fermions (graphene) is the analogue of the well known Zeeman term for non-relativistic massive particles as can be learned from Sakurai \cite{Sakurai2010}. Briefly, one branch of the squared Dirac equation for massive fermions reads:
\begin{equation}
\left[ \vec{\sigma} (\vec{p} -\vec{{\cal{A}}}) \right]^2 u = (E^2 - m^2) u
\label{eq:8},
\end{equation}
we use that $E^2 - m^2$ is the momentum squared $p^2$. In the non relativistic limit $p^2 = 2E_{\rm kin} m$ ($E_{\rm kin}$: kinetic energy), leading to:
\begin{equation}
\frac{1}{2m} \left[ \vec{\sigma} (\vec{p} -\vec{{\cal{A}}}) \right]^2 u = E_{\rm kin} u
\label{eq:9}
\end{equation}
The solution to $(\vec{\sigma} (\vec{p} -\vec{{\cal{A}}}))^2$ we have seen previously (eq. \ref{eq.:H2LL}), so we find:
\begin{equation}
\left[ \frac{(\vec{p} -\vec{{\cal{A}}})^2}{2m} - \vec{\mu} \vec{B} \right] u = E_{\rm kin} u
\label{eq:massivezeeman}
\end{equation}
where $\vec{\mu} = \frac{e}{m} \vec{S}$ and $\vec{S} = \frac{\hbar}{2} \vec{\sigma}$. This derivation shows that the familiar Zeeman term $\vec{\mu} \vec{B}$ appears by squaring the Dirac Hamiltonian for massive particles in the non-relativistic limit.

Further insight into the pseudo-Zeeman term can be gained, if we consider the situation where the kinetic energy is quantized within a homogeneous $B_{\rm ps}$.
In this case, the pseudo-Zeeman term exactly cancels the energy of the lowest cyclotron orbit for sublattice A  \cite{Katsnelson2012}.
This leads to a fully pseudospin polarized Landau level at zero energy, a hallmark of graphene \cite{Settnes2016, Katsnelson2012}. In the case of inhomogeneous $B_{\rm ps}$ it also leads to a sublattice polarization \cite{Wehling2008a, Katsnelson2012}.

\refstepcounter{supplement}
\section*{S\arabic{supplement}: Molecular dynamics calculations}
\addcontentsline{toc}{section}{S\arabic{supplement}: Molecular dynamics calculations}
\label{sec:md}
\refstepcounter{alter}
\subsection*{S\arabic{supplement}-\arabic{alter}: Van der Waals interaction between tip, graphene and SiO$_2$}
\addcontentsline{toc}{subsection}{S\arabic{supplement}-\arabic{alter}: Van der Waals interaction between tip, graphene and SiO$_2$}
\label{sec:mdvdw}

The lifting of graphene by the W tip can be rationalized by considering the polarizabilities of the contributing atoms.
The polarizabilities according to Hartree-Fock calculations for the free, neutral atoms of SiO$_2$, C and W are \cite{fraga1976handbook}: $\alpha_{\mathrm{W}}$ = 21.4 \AA$^3$, $\alpha_{\mathrm{Si}}$ = 6.81 \AA$^3$, $\alpha_{\mathrm{O}}$ = 0.73 \AA$^3$, $\alpha_{\mathrm{C}}$ = 1.74 \AA$^3$.
Since the van der Waals (vdW) potential is proportional to $\alpha^2$, the attractive force between a distant tungsten tip and graphene can be larger than the vdW force pinning the graphene to the surface. 
According to ab initio calculations \cite{Wolloch2015}, even graphene on Ir(111) ($\alpha_{\mathrm{Ir}}$ = 15.6 \AA$^3$) can be lifted by a W tip by up to 0.5 \AA.

To model the lifting in our experiments, molecular dynamics simulations have been performed using the LAMMPS code \cite{Plimpton1995}. Therefore, the interactions between the tungsten tip, the graphene and the SiO$_2$ have been modelled using a pairwise Lennard-Jones (LJ) potential of the form: $V_{\mathrm{LJ}} = 4 \varepsilon \left[ \left( \frac{\sigma}{r} \right)^{12} - \left( \frac{\sigma}{r} \right)^6 \right]$.

Parameters for the Si-Si, the O-O, and the C-C interaction are taken from the universal force field (UFF) \cite{Rappe1992}.
Parameters for the tip-graphene and graphene-substrate interactions are generated by Lorentz-Berthelot mixing rules, for example:
\begin{equation}
\varepsilon_{\mathrm{C-Si}} = \sqrt{\varepsilon_{\mathrm{C-C}} \varepsilon_{\mathrm{Si-Si}}} \text{\ \ \ and\ \ \ }
\sigma_{\mathrm{C-Si}} = \frac{\sigma_{\mathrm{C-C}} + \sigma_{\mathrm{Si-Si}}}{2}
\label{eq:21}
\end{equation}
This is not possible for the W atoms, since the UFF parameters refer to the cationic state of the metal.
Therefore, the polarizability of W adatoms on a W (110) tip, as measured by field ion microscopy experiments \cite{Wang1982, Tsong1971} was used as $\alpha_{\mathrm{W}}$.
These polarizability values are considered to be a good approximation of our experimental system, since the most likely STM tip orientation is a (110) pyramid \cite{Yerra2005}.
Using $\alpha_{\mathrm{W}}$, the $C_6$ coefficient of the van der Waals potential $(V_{\mathrm{vdW}} = -C_6/r^6)$ for the C-W interaction was determined by the Slater-Kirkwood formula \cite{Bichoutskaia2008}.
The C-W LJ parameters ($\varepsilon_{\mathrm{C-W}}$, $\sigma_{\mathrm{C-W}}$) were determined by fitting the attractive part of the 12-6 LJ curve to the $C_6/r^6$ potential. Due to the uncertainties in the experimental polarizability, an upper and a lower bound for the LJ parameters were used.
In both cases lifting of graphene supported by SiO$_2$ was found by the simulations.
\begin{table}
\begin{center}
\begin{tabular}{ l | c | r }
& $\varepsilon$ [meV] (min, max) & $\sigma$ [\AA] \\
\hline
C-O interaction & 3.442 & 3.27 \\
C-Si interaction & 8.909 & 3.62 \\
C-W interaction & 65, \textbf{120} & 3.2 \\
O-W interaction & 9.6, \textbf{13} & 3.16 \\
Si-W interaction & 104, \textbf{142} & 3.51 \\
\end{tabular}
\caption{Lennard-Jones parameters. Bold values are used for the calculations shown in supplementary Fig. \ref{fig:s3} and Fig. 3a-c of the main text}
\label{table:s1}
\end{center}
\end{table}
Using these LJ parameters, the adsorption energy of graphene on amorphous SiO$_2$ has been calculated, resulting in a value of 43.53 meV/atom.
This agrees well with the adsorption energy calculated from first principles, with values between 32.4 meV/atom and 55.1 meV/atom (DFT with dispersion corrections) \cite{Gao2014}, as well as with measurements of the adhesion (56.7$\pm$2.6 meV/atom) \cite{Na2014}.

Note that we must differentiate between the tip sample separations in the calculations $z^*$ and the tunnelling distance $z$.
The quantity $z^*$ differs from $z$ by an offset, since $z$ = 0 \AA\ is taken as the distance where the conductance between sample and tip reaches the conductance quantum \cite{Kroger2007} $G_0 = 2 e^2/h$.
However, the exact offset of $z$ with respect to $z^*$ could only be deduced by detailed transport calculations using a tip with known atomic configuration.
Generally, $z^*$ delicately depends on the tunnelling orbitals and their respective vdW-radii.
We assume $z^*-z$ to be the sum of the vdW-radius of tungsten and the length of the p$_z$-orbital of graphene, i.e. $z^*-z = 2-4$ \AA.

\refstepcounter{alter}
\subsection*{S\arabic{supplement}-\arabic{alter}: Details of the LAMMPS calculation}
\addcontentsline{toc}{subsection}{S\arabic{supplement}-\arabic{alter}: Details of the LAMMPS calculation}
\label{sec:lammps}

The bonds in between the tungsten atoms was implemented via the embedded-atom method potential \cite{Zhou2001}, while the Tersoff potential was used for the SiO$_2$ substrate \cite{Munetoh2007}.
For graphene we used the AIREBO potential \cite{Stuart2000}.
Visualization of the data was done using OVITO \cite{Stukowski2010}.
The calculations were performed for a cell of size (84.6 $\times$ 83) \AA$^2$ in the $(x,y)$-plane and 70 \AA\ in the $z$ direction.
For this cell size, the graphene is strain-free due to the fitting of the periodic boundary conditions used in the $(x,y)$-plane to the atomic lattice.
The tip is modelled as a pyramid made up of stacked W(110) planes.
The positions of the atoms in the top layer of the tip were fixed, while the rest of the tip could relax during energy minimization.
Fixing the whole tip or letting it relax during the simulation does not affect the height of the graphene deformation.
The experimentally most likely situation involves a tip, which does not show any rotational symmetry along the $z$ axis, due to a misalignment of the (110) planes of the tip with respect to the W wire axis or due to a rotation of the tip with respect to the sample plane.
To model this situation in the molecular dynamics calculations, the (110) crystallographic planes of the STM tip are tilted by 30$^\circ$ with respect to the graphene surface towards the zig-zag direction of graphene.

\begin{figure}[h]
\includegraphics[width = 1 \textwidth]{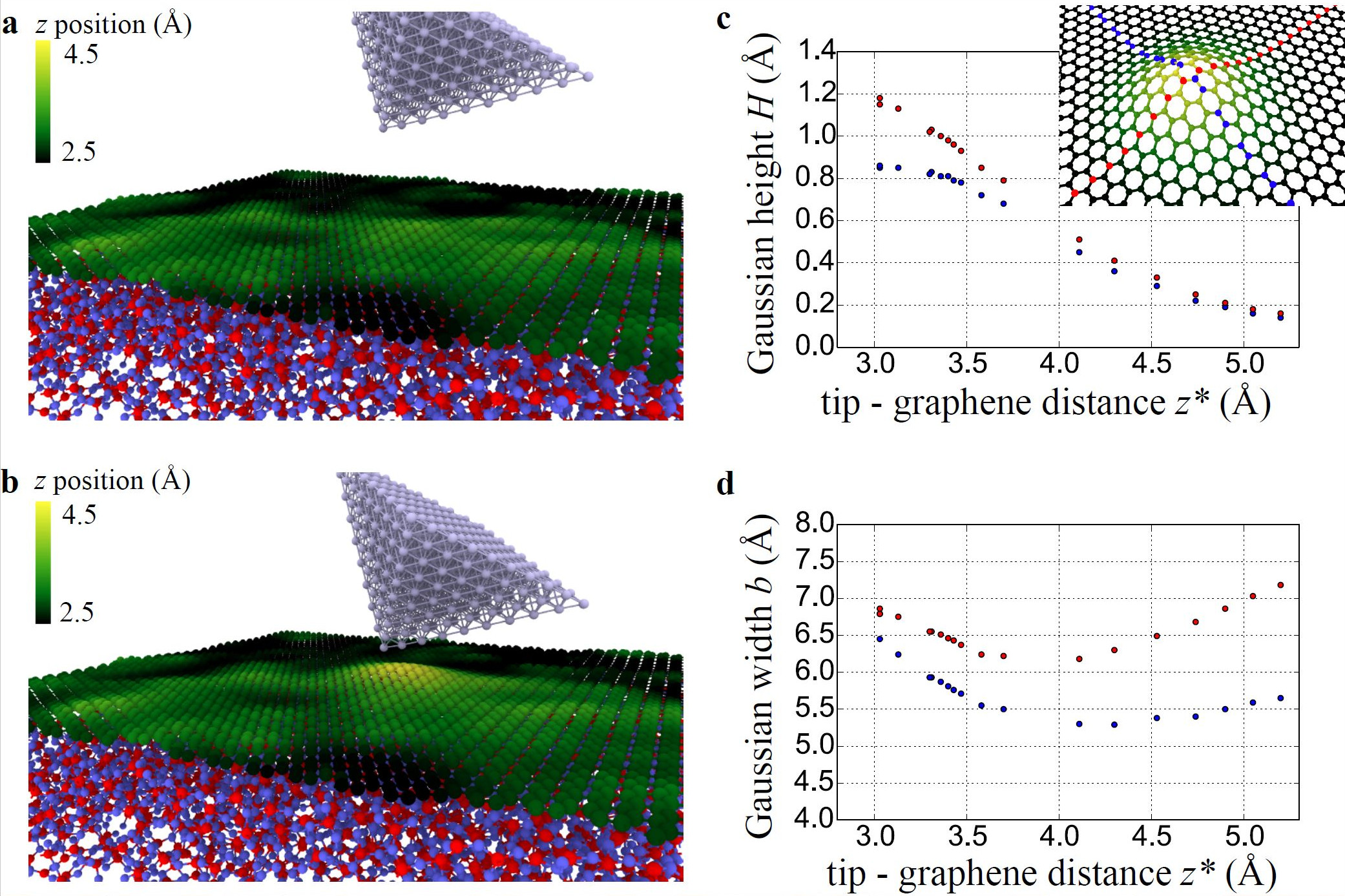}
\caption{\textbf{LAMMPS simulation.} \textbf{a}, Ball model of graphene on SiO$_2$ with the tip far away from graphene ($z^*$ = 7 \AA) (W atoms: grey, C atoms: green, O atoms: blue, Si atoms: red).
Colour code on graphene indicates the local height with respect to the SiO$_2$. The amorphous SiO$_2$ surface induces a slight corrugation of the graphene. \textbf{b}, Same as (a) at $z^*$ = 3.5 \AA.
Moving the tip close to the graphene induces a Gaussian deformation of height $H \simeq$1 \AA. \textbf{c, d}, Height $H$ and width $b$ of the Gaussian deformation for various tip-graphene distances $z^*$ as obtained by fitting a Gaussian curve to the graphene atomic positions along the armchair (blue) and zigzag (red) directions (see inset). Red marks are along the tilt direction of the tip. Notice that the deformation becomes more asymmetric at $z^*\le 3.7$ \AA\ as visible by the different $H$ of the Gaussian fits in perpendicular directions.
The difference between $z^*$ (distance between atom cores) and $z$ (the distance where tunnelling conductance is $2e^2/h$) is described in Sec. S\ref*{sec:md}-\ref{sec:mdvdw}}
\label{fig:s3}
\end{figure}

The SiO$_2$-substrate is prepared by annealing $\alpha$-quartz in a periodic simulation cell, at 6000 K with a time step of 0.1 fs.
The system was held at 6000 K for 10 ps, after which the temperature was lowered to 300 K at a rate of $10^{12}$ K/s, over 570 ps \cite{Ong2010}.
After quenching the system, the energy was minimized via the conjugate gradient method.
The radial distribution function of the amorphous SiO$_2$ obtained in this way matches that of SiO$_2$ glass as known from x-ray data \cite{Warren1934}.
To prepare the SiO$_2$ surface, the atoms in the top half of the amorphous SiO$_2$ are removed and the surface is subsequently relaxed. 
Alternatively, a simplified substrate has been used in the calculations.
This substrate is modelled by a featureless surface 3.09 \AA\ beneath the graphene, acting on the graphene atoms with a force perpendicular to the surface.
For this "wall" type substrate, we used the "9-3" Lennard-Jones potential of the form: $V_{\mathrm{LJ}} = \varepsilon \left[ \frac {2}{15} \left(\frac{\sigma}{r}\right)^9 - \left( \frac{\sigma}{r} \right)^3 \right]$, which describes the vdW interaction between a surface and an atom \cite{Israelachvili}.
We have chosen the $\varepsilon$ and $\sigma$ parameters such that the graphene adsorption energy on this substrate is 43.53 meV/atom.
The choice of either the amorphous SiO$_2$ or the wall-type substrate does not have any influence on the height and width of the graphene deformations produced by the tip.
Therefore, in most calculations the wall type potential is used in order to save computational time and to avoid the slight graphene corrugations induced by the amorphous SiO$_2$ surface (\ref{fig:s3}a).
Lifting of graphene was obtained by relaxing the graphene-substrate-tip system via energy minimization with the tip far away from the surface, followed by lowering the tip towards the graphene and running another energy minimization via the conjugate gradient method.
The simulation results in graphene deformations with up to $H \approx$ 1 \AA\ indicating that graphene can be lifted by the tip, if originally in contact with SiO$_2$.

\begin{figure}[h]
\includegraphics[width = 1 \textwidth]{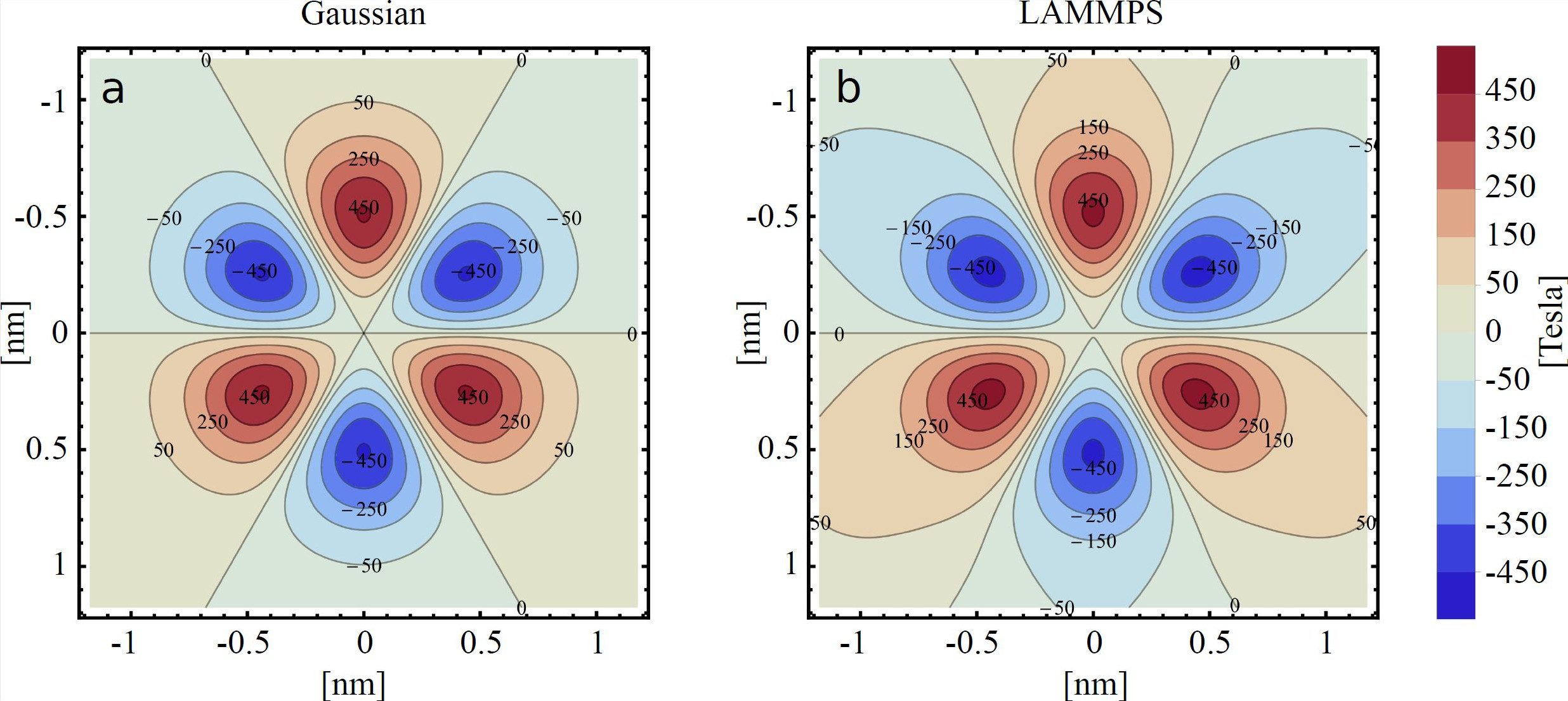}
\caption{\textbf{Contour plot of $B_{\rm ps}$ in Gaussian/LAMMPS deformation:} \textbf{a}, $B_{\rm ps}$ of a Gaussian deformation with $H$ = 1 \AA\ and width $b$ = 5.8 \AA. \textbf{b}, $B_{\rm ps}$ of a deformation from LAMMPS calculations, which is fitted by a Gaussian with parameters: $H$ = 1 \AA, $b$ = 5.8 \AA.}
\label{fig:gau_lammps}
\end{figure}

The deformations resulting from the LAMMPS calculations are well fitted by a Gaussian of the form: $h(r) = H \cdot \exp(-r^2/b^2)$, as shown in Fig. 3a-c of the main text.
Therefore, within our tight binding and continuum Dirac model calculations, we have used this Gaussian function to describe the displacement of the graphene membrane.
The major difference between the LAMMPS and Gaussian deformation is that the one resulting from molecular dynamics will have in plane relaxation of the atoms, in addition to the out of plane displacement.
To check the validity of our Gaussian approximation we compare the pseudo-magnetic field of a deformation resulting from LAMMPS calculations with a perfect Gaussian.
For the latter, we fit the LAMMPS deformation we determine the height $H$ and width $b$ and calculate $B_{\mathrm{ps}}$, according to Eq (3) from ref. \cite{Schneider2015}.
The strain tensor of the LAMMPS deformation was evaluated by fitting algebraic functions to the in plane and out of plane atom displacements.
\ref{fig:gau_lammps} shows the comparison of $B_{\rm ps}$ for the perfect Gaussian and the LAMMPS deformation.
The maximum $B_{\rm ps}$ difference is 14\% or 60 T, 1.2 nm or $\sim$2$b$ away from the deformation maximum.
The $B_{\rm ps}$ distribution of the LAMMPS deformation is only slightly asymmetric, reflecting the $C_2$ symmetry of the (110) W STM tip, which is barely visible in \ref{fig:gau_lammps}b.
These calculations show that the Gaussian approximation used in the main text (e.g. Fig 3D) is valid.

\newpage

\refstepcounter{alter}
\subsection*{S\arabic{supplement}-\arabic{alter}: Estimating the energy contributions during graphene lifting}
\label{sec:energies}
\addcontentsline{toc}{subsection}{S\arabic{supplement}-\arabic{alter}: Estimating the energy contributions during graphene lifting}

Here, we disentangle the different interaction forces in order to get a more intuitive understanding of the observed lifting.

\begin{figure}[h]
\includegraphics[width = 1 \textwidth]{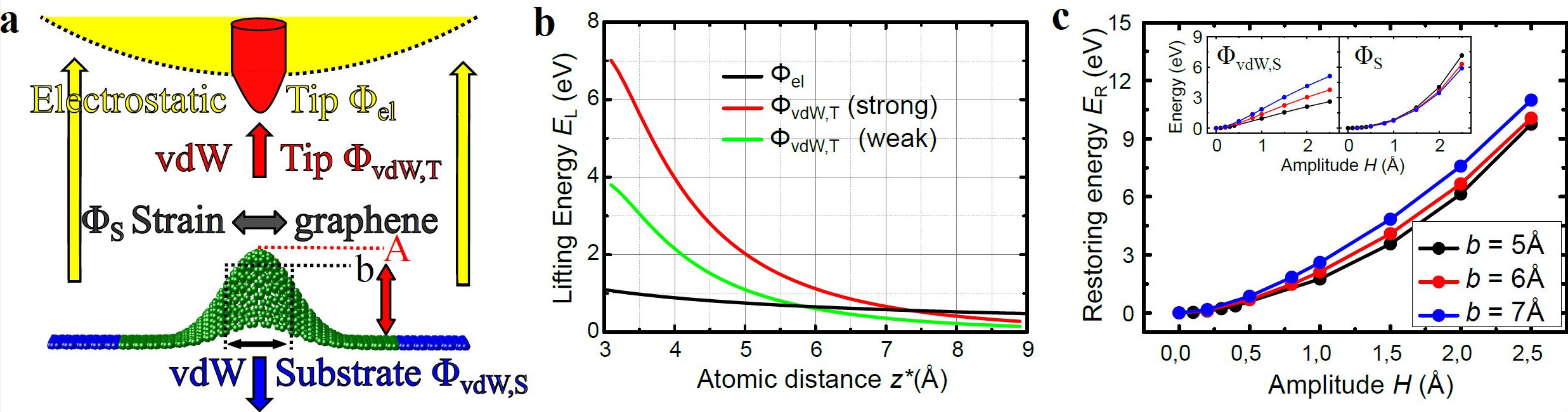}
\caption{\textbf{Energy scales of the lifting process.} \textbf{a}, Schematic of the forces involved in the lifting process. Arrows labelled with the associated potential energies $\Phi$ and the origin of the force (tip, substrate, graphene) mark the direction of the forces acting on graphene. The yellow half-sphere represents the large-scale part of the tip responsible for the electrostatic potential $\Phi_{\rm el}$. Lifting amplitude $H$ and width $b$ of the Gaussian deformation are marked as resulting from the MD. \textbf{b}, Potential energies, which favour the lifting of graphene $E_{\rm L}$. $z^*$ is the vertical distance between the cores of the closest atoms of the tip and graphene. The vdW-potential $\Phi_{\rm vdW, T}$ is calculated by LAMMPS using a pairwise model between a 30$^o$ tilted W(110) tip and a (12 nm)$^2$ area of graphene. Large (red) or small (green) polarizabilities of the W atoms are applied according to Table \ref{table:s1}. $\Phi_{\rm el}$ (black line) is calculated in a sphere-plane geometry for tunnelling voltage $V$ = 0.5 V \cite{Mashoff2010}. \textbf{c}, Sum of potential energies, which oppose the lifting ($E_{\rm R}$) as a function of the amplitude of the Gaussian deformation  for different widths of the Gaussian $b$ calculated in the absence of the tip. Left inset: vdW-potential between graphene and SiO$_2$ ($\Phi_{\rm vdW, S}$). Right inset: strain potential of the deformation $\Phi_{\rm S}$ calculated using the AIREBO potential \cite{Stuart2000}. Both potentials are calculated by LAMMPS \cite{Plimpton1995}.}
\label{fig:s4}
\end{figure}

The interaction forces are sketched in Fig. \ref{fig:s4}a. The force directions are marked by coloured arrows.
The potential energy $E_{\rm L}$ favouring the lifting is the sum of the vdW potential $\Phi_{\rm vdW, T}$ between tip and graphene and the electrostatic energy $\Phi_{\rm el}$ caused by the differences between the electrostatic potentials of tip and sample \cite{Mashoff2010}.
The restoring potential energy $E_{\rm R}$, opposing the lifting, is the sum of the vdW potential between graphene and substrate $\Phi_{\rm vdW, S}$ and the strain potential within the Gaussian deformation $\Phi_{\rm S}$ (Fig. \ref{fig:s4}a).
In Fig. \ref{fig:s4}b and c, we plot these energies as a function of $z^*$, the distance between the atomic cores of the atoms of tip and graphene being closest to each other.
Since the tunnelling distance $z$ can be calculated by Eq. \ref{eq:17}, we get, e.g. for  $I$ = 50 nA at $V$ = 0.5 V, $z \approx$ 2 \AA\ corresponding to $z^* \approx 4-6$ \AA.
$\Phi_{\rm vdW}$ is then calculated in the pairwise model as described in S\ref*{sec:md}-\ref{sec:mdvdw}, using the upper and lower values for the polarizabilities as found in the literature (Table \ref{table:s1}). Results for the Gaussian deformation found by the MD simulations of a pyramidal tungsten tip tilted by 30$^o$ above a circular graphene area of (12 nm)$^2$ are displayed in Fig. \ref{fig:s4}b. For $z^*$ = 4.5 \AA, e.g., we find $\Phi_{\rm vdW, T}$ = 1.5-3 eV, which changes only slightly if other reasonable tip geometries are used.

The electrostatic energy $\Phi_{\rm el}$ between tip and graphene is estimated by $\Phi_{\rm el} = \frac{1}{2} \displaystyle{\cal{C}}V^2$, where $\cal{C}$ is the capacitance of the system.
We calculated $\cal{C}$ following a model described elsewhere \cite{Mashoff2010}.
In short, the tip is represented by a W sphere of radius 5 nm and the sample as a circular plate of graphene with radius 2 nm, taking into account the finite charge carrier density of graphene (quantum capacitance).
The resulting $\Phi_{\rm el}$ is smaller than $\Phi_{\rm vdW, T}$ for $z^* < 7$ \AA, i.e. for all reasonable distances during lifting.
Nevertheless, it provides an approximately constant background energy of 0.5-1 eV favouring lifting.
Since the electrostatic forces are more homogeneous than the vdW-forces, the deformation shape of graphene will, however, be dominated by the stronger and more short-range vdW-potentials.

The sum of the restoring potentials $\Phi_{\rm vdW, S} + \Phi_{\rm S}$ is plotted for different Gaussian deformation geometries in Fig. \ref{fig:s4}c.
The two contributing potentials $\Phi_{\rm vdW, S}$ and $\Phi_{\rm S}$ are plotted separately in the inset.
$\Phi_{\rm vdW, S}$ is calculated in the absence of the tip by applying the wall-type potential between graphene and SiO$_2$, fitting to the experimental adhesion energy as described in S\ref*{sec:md}-\ref{sec:lammps}.
$\Phi_{\rm S}$ is simulated using the AIREBO potential \cite{Stuart2000} of graphene. The comparison of Fig. \ref{fig:s4}b and c reveals that, e.g., at $z^* \approx$ 4.5 \AA\ ($z \approx$ 2 \AA, $I \approx$ 50 nA), the lifting energies $\Phi_{\rm vdW,T} + \Phi_{\rm el} \approx$ 2.5-4 eV can induce a Gaussian amplitude of $H$ = 1-1.5 \AA.
This reasonably agrees with the lifting heights found in the MD and with the lower experimental lifting heights presumably found in supported areas of graphene (Fig. \ref{fig:s2}d, and Fig. 2d, f of the main text, blue areas).
Thus, we corroborate that lifting by the STM tip can also appear on supported graphene within tunnelling distance. 

In turn, lifting heights of $H_{\rm exp} \approx$ 2.5-3 \AA, as partly observed in the experiment, are not possible in tunnelling distance according to our estimates.
Thus, they can only be realized for areas that are originally not in contact with the substrate, such that $\Phi_{\rm vdW,S}$ is significantly reduced.
Such areas have indeed been found previously for graphene on SiO$_2$ \cite{Mashoff2010, Geringer2009}.
In those areas, mostly the strain energy $\Phi_{\rm S}$ has to be paid for the lifting allowing larger amplitudes.
For example, for $z^*=4.5$ \AA\ (50 nA, 0.5 V) without $\Phi_{\rm vdW, S}$, we find $H = 1.8-2.2$ \AA\ again in reasonable agreement with the experiment.

\refstepcounter{alter}
\subsection*{S\arabic{supplement}-\arabic{alter}: Video - Movement of deformation with scanning tip}
\addcontentsline{toc}{subsection}{S\arabic{supplement}-\arabic{alter}: Video - Movement of deformation with scanning tip}

To create an animation of the moving deformation during scanning of the STM tip, we have taken advantage of the fact that the graphene is periodic within the calculation cell.
After the energy minimization, the graphene was laterally moved by 0.141 \AA\ and the energy of the system was minimized again. By repeating this step, the scanning of the STM tip was simulated. After the tip has travelled one graphene unit cell, the movie is looped. LJ parameters are the same as in the other MD calculations.

% ------------------------------------------------------------------

\refstepcounter{supplement}
\section*{S\arabic{supplement}: Excluding alternative models that predict sublattice symmetry breaking}
\addcontentsline{toc}{section}{S\arabic{supplement}: Excluding alternative models that predict sublattice symmetry breaking}

In order to substantiate our successful description of the SSB by pseudospin polarization, we have to exclude other possible mechanisms. In the following sub-chapters we consider:
\begin{enumerate}
\item \textbf{The influence of double or multiple tunnelling tips.}\\
\emph{The dependence of the SSB on the tunnelling current and hence the lifting height, would be opposite.}

\item \textbf{A different lifting height of the graphene membrane, if the centre of the tip is positioned either on sublattice A or on sublattice B.}\\
\emph{The effect is at least a factor of 100 too small to explain the SSB.}

\item \textbf{Real buckling of the graphene lattice as present, e.g., in silicene \cite{Cahangirov2009}.}\\
\emph{It requires a compressive strain of 16\%, which is of the wrong sign (pulling graphene implies tensile strain) and a factor of, at least, 10 smaller than the applied strains.}

\item \textbf{A Peierls transition as expected to be possible in graphene due to the Kohn anomaly and other types of Kekul\'{e} order.}\\
\emph{It requires an expansion of the graphene lattice by 12\%, which is again a factor of 10 too large (see S\ref{sec:energies}).}

\item \textbf{A sublattice symmetry breaking due to the correlation of electric and pseudo-magnetic fields as proposed by Low, Guinea and Katsnelson \cite{Low2011}.}\\
\emph{The SSB should be voltage dependent, which we don't observe up to 1 V. Moreover, it is likely a factor of 10 smaller than the observed SSB.}
\end{enumerate}
Consequently, these models fail either qualitatively or quantitatively by a large margin when compared with our experimental results, which we describe in detail in the following.

\refstepcounter{alter}
\subsection*{S\arabic{supplement}-\arabic{alter}: Multiple tips}
\addcontentsline{toc}{subsection}{S\arabic{supplement}-\arabic{alter}: Multiple tips}

It is well known that STM images are prone to artefacts arising from multiple tips contributing to the tunnelling current.
Multiple tip effects can generally be ordered in two categories: either the two scanning tips are far from each other or they are close on the scale of the Bloch function periodicity of the sample wave functions.
In the first case \cite{Mizes1987}, the contributions of the two tunnelling tips sum up, resulting in "ghost images" from the secondary tip.
In the second case \cite{Mizes1987}, interference can occur between the two tunnelling channels \cite{DaSilvaNeto2013}.
This induces a symmetry breaking within the STM images, reflecting the rotational symmetry of the tip.

In the former case, one could imagine that two graphene lattices imaged by two different tips are overlaid in a way that sublattice A imaged by tip 1 overlaps with sublattice B imaged by tip 2, while sublattice B by tip 1 does not overlap with sublattice A by tip 2.
This would lead to an apparent SSB and a weaker additional spot within the graphene unit cell belonging to sublattice A imaged by tip 2.
Firstly, we never observe such an additional spot, if we see SSB.
Secondly, one would expect that double tips are less important, if one moves the mostly imaged area towards the tip, thereby enhancing its contribution to the image with respect to the ghost image.
Consequently, the SSB should disappear with increased lifting height in striking contrast to the experimental finding.
Thus, we can safely rule out long-range double tips as the origin of SSB.

If the two tunnelling tips are close together on the scale of the Bloch function wavelength, interference effects can arise between sample quasiparticle states and those of the tip \cite{DaSilvaNeto2013}.
In this case, the rotational asymmetry of the STM tip is transferred to the STM images.
In order to create a sublattice symmetry breaking, the tip would need to have threefold symmetry $C_3$ (see Fig. \ref{fig:s8}). 
Such interference effects are typically strongly energy dependent, such that they only appear in differential conductance maps \cite{DaSilvaNeto2013}.
Firstly, we measure topography images at relatively large voltage ($V$ up to 1V), i.e., we integrate over the various interference terms.
As shown by da Silva Neto et al. \cite{DaSilvaNeto2013}, this results in overall cancellation of the asymmetries.
Secondly, we do not see any drastic changes in the SSB pattern (disappearance and reappearance \cite{DaSilvaNeto2013}) between $V=0.05$ V and $V=1$ V.
Thirdly, it is highly unlikely that the tip asymmetry causes strongly different strength of the interference on originally supported and suspended areas independent of the lateral shape of the corresponding areas.
Thus, we rule out this possibility.

\refstepcounter{alter}
\subsection*{S\arabic{supplement}-\arabic{alter}: Tip induced favoured lifting}
\addcontentsline{toc}{subsection}{S\arabic{supplement}-\arabic{alter}: Tip induced favoured lifting}

\begin{figure}[h!]
\includegraphics[width = 1 \textwidth]{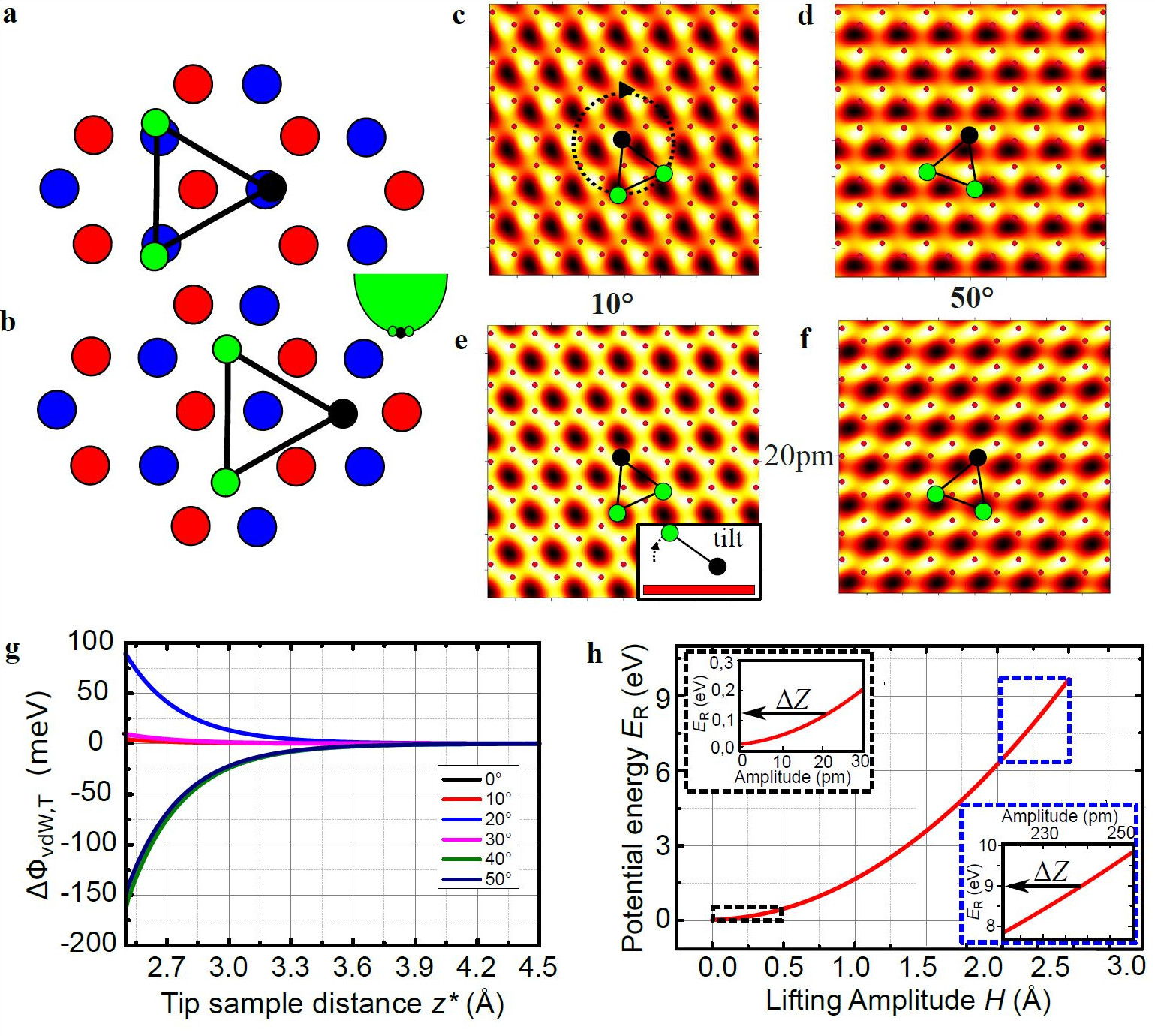}
\caption{\textbf{Favoured lifting above one sublattice.} \textbf{a}, \textbf{b}, Schematic of an STM tip consisting of 3 atoms (green and black dots) with the centre located (\textbf{a}) on sublattice A (red dots) or (\textbf{b}) on sublattice B (blue dots). Note that the three tip atoms are either on top of C atoms (a) or not (b). Inset: Sketch of the tip apex (green) with the three terminating tip atoms marked. The atom closest to graphene which dominates the tunnelling current is coloured black. \textbf{c}$-$\textbf{f}, The vdW-potential energy $\Phi_{\rm vdW, T}(x,y)$ between the three tip atoms and graphene as a function of lateral tip position using an unrealistically small tip sample-distance $z^*$ = 2 \AA\ (distance between centres of black tip atom and graphene plane). Atomic positions of graphene are marked by red dots. (c), (d) All three tip atoms have the same distance to the graphene plane. (e), (f) Green tip atoms are 20 pm further apart from the graphene plane. The azimuthal angle of the tip with respect to the armchair direction is marked: (c), (e) $\varphi$ = 10$^o$, (d), (f) $\varphi$ = 50$^o$. The tip with the black atom being the closest to graphene is sketched in all images and the tilted tip is visualized in the inset of (e). \textbf{g}, Potential difference $\Delta \Phi_{\rm vdW, T, (A-B)}$ between the two sublattices as deduced from images as (c)$-$(f). Different tip azimuths $\varphi$ are labelled. \textbf{h}, Required potential energy $E_{\rm R}$ to induce a Gaussian deformation ($b$ = 6 \AA) of graphene on SiO$_2$ according to molecular dynamics (see Fig. \ref{fig:s4}). Insets show zooms into the region of negligible (large) lifting amplitudes, marked by black (blue) squares. Arrows mark the energy required for an additional lifting of graphene by $\Delta z$ = 20 pm.}
\label{fig:s8}
\end{figure}

Another possibility is that the apparent sublattice height difference $\Delta z$ is induced by a different lifting amplitude  $H_{\rm exp}$, if the tip is positioned either on sublattice A or on sublattice B.
This requires an anisotropic distribution of vdW forces.
Figures \ref{fig:s8}a and b show the last three tip atoms (green and black) within a maximally anisotropic triangular configuration centred either atop sublattice A (a) or sublattice B (b).
The three tip atoms if centred atop sublattice A or B are located atop graphene atoms or atop holes of the hexagons, respectively.
This naturally leads to different vdW forces for these two cases.
We neglect the more slowly decaying electrostatic potentials $\Phi_{\rm el} \sim 1/r$, since their variation on the atomic scale (C-C distance $<$ tip graphene distance) is negligible with respect to the one from the more short-range vdW potentials  $\Phi_{\rm vdW} \sim 1/r^6$.
In order to quantify the difference in lifting, we simulate the local pairwise vdW-potentials $\Phi_{\rm vdW}$ of three atoms of a W tip, which form a W(110) facet with corresponding inter-atomic distances, with the C atoms of a (12 nm)$^2$ area of graphene.
We only use the attractive part of the Slater-Kirkwood formula \cite{Bichoutskaia2008}.
Simulations including a second layer of tip atoms above the triangle reveal that the additional atoms do not alter the differences of vdW forces on the atomic scale.
Adding a single W atom to the graphene side of the tip triangle moves the triangle so far apart from graphene that the differences of vdW forces on the atomic scale are suppressed by more than one order of magnitude.
Thus, for the sake of simplicity, we consider a single triangle of tip atoms with the strongest possible differences in vdW forces. 

The pairwise interactions are calculated using the polarizabilities of Table \ref{table:s1} and are subsequently summed up to reveal $\Phi_{\rm vdW ,T}$.
The scanning of the tip is simulated by moving the three tip atoms laterally on the graphene lattice at a constant tip sample distance $z^*$, which is the vertical distance between the centres of the last tip atom and the closest C atom of graphene.
$z^*$ is larger than the tunnelling distance $z$ between tip and graphene by $2-4$ \AA\ (see S\ref*{sec:md}-\ref{sec:mdvdw}), since $z=0$ \AA\ is taken to be at tunnelling conductivity $\sigma = 2e^2/h$.
Figure \ref{fig:s8}c-f shows the resulting scanned  $\Phi_{\rm vdW, T}(x,y)$ at an unrealistically small $z^*$ = 2 \AA, which is artificially possible since we ignore chemical bonding forces.
This results in a relatively strong SSB.
We used different azimuthal angles $\varphi$ of the tip with respect to the armchair direction of graphene as sketched in all images and different vertical tilts of the tip as sketched in Fig. \ref{fig:s8}e.
For some $\varphi$, we find potential patterns that break the sublattice symmetry (Fig. \ref{fig:s8}c-f). Figure \ref{fig:s8}g shows the potential difference between the two sublattices $\Delta \Phi_{\rm vdW, T, (A-B)} = \Phi_{\rm vdW,T,(A)} - \Phi_{\rm vdW ,T,(B)}$ for different $\varphi$ and $z^*$ showing that $\Delta \Phi_{\rm vdW, T, (A-B)}$ strongly decreases with increasing $z^*$.
For the estimated distances at largest tunnelling current $I$ = 50 nA being $z^* \approx$ 4-6 \AA\, the energy difference is $\Delta \Phi_{\rm vdW, T, (A-B)} \leq 1$ meV.

This can be compared with the energy cost for lifting.
For small $z^*$, we have to take into account that the graphene is already lifted by $H \approx$ 1.5 \AA\ (Section S\ref*{sec:md}-\ref{sec:energies}).
The energy cost to increase $H$ by an additional  $\Delta z$ = 0.2 \AA, as observed experimentally, is  $\sim$1 eV according to the MD simulations (Fig. \ref{fig:s8}h), i.e. more than three orders of magnitude larger than $\Delta \Phi_{\rm vdW, T, (A-B)}$.
Even, if we assume that for an unknown reason, graphene is not lifted at all, if the tip is positioned on one sublattice, the required cost to lift the graphene with a tip on top of the other sublattice would be $\sim$0.1 eV still two orders of magnitude too large. 

For a consistent model, one has to additionally allow imaging with atomic resolution, which is not provided by a planar triangle of tip atoms.
Tilting the tip as shown in Fig. \ref{fig:s8}e-f, however, reduces $\Delta \Phi_{\rm vdW, T}$ further.
These quantitative estimates safely exclude the scenario of a favoured lifting with the tip centred above one of the sublattices as an explanation of the observed SSB.

\refstepcounter{alter}
\subsection*{S\arabic{supplement}-\arabic{alter}: Compression induced buckling}
\addcontentsline{toc}{subsection}{S\arabic{supplement}-\arabic{alter}: Compression induced buckling}

\begin{figure}
\includegraphics[width = 1 \textwidth]{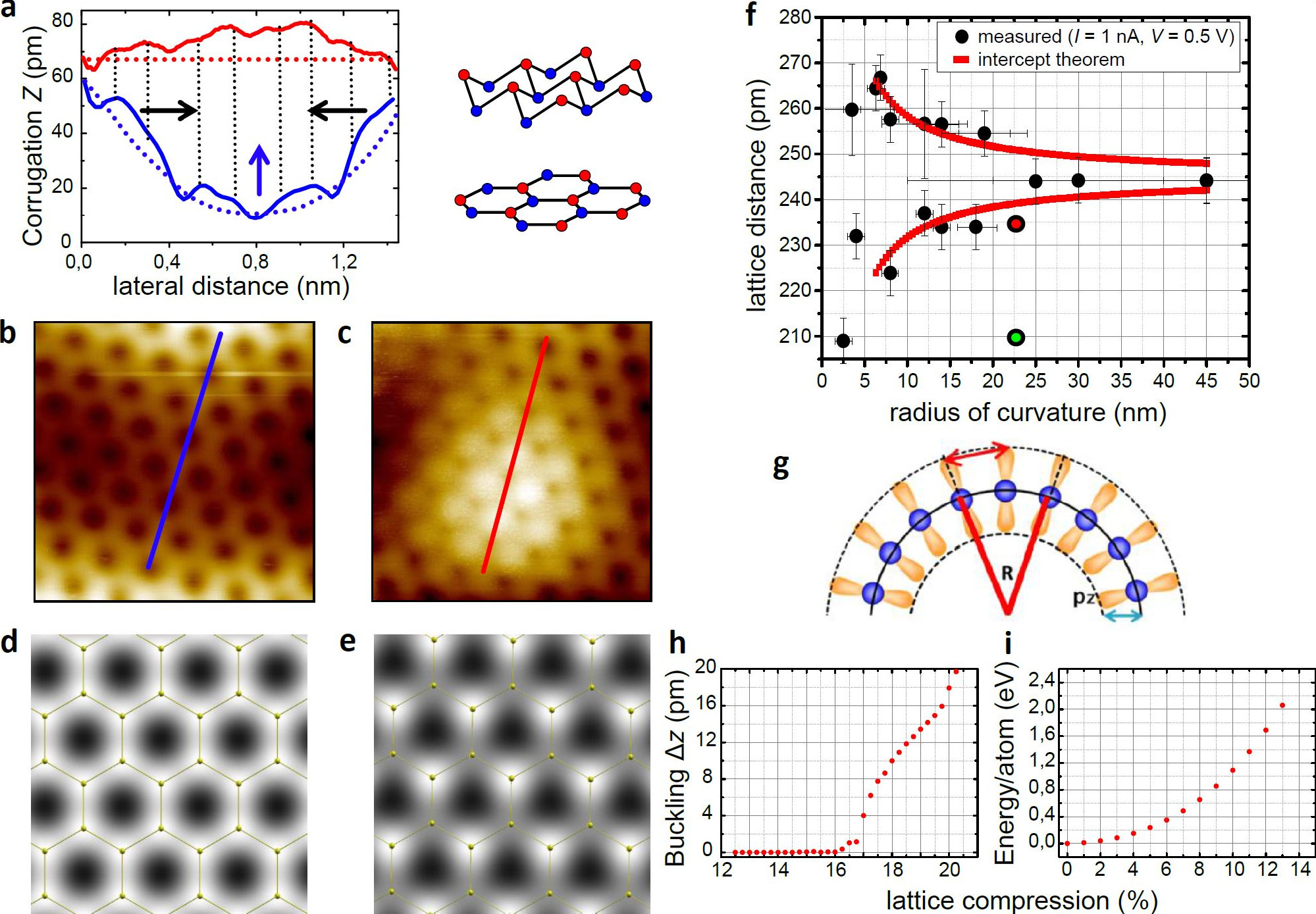}
\caption{\textbf{Compression induced buckling.} \textbf{a}, Profile lines for relaxed (blue) and lifted (red) graphene as indicated in (b).
Dotted lines (red, blue) mark the low pass topography of the graphene (see S\ref{sec:sublcontr}); graphene lifting (blue arrow) and compression (black arrows) is indicated; same atoms are marked by dotted black lines.
Modifications of the graphene lattice are illustrated for buckled (top) and flat (bottom) graphene. Blue and red dots mark different sublattices.
\textbf{b}, \textbf{c} STM image of the same (1.5 nm)$^2$ of graphene (b) in the relaxed ($I$ = 0.1 nA, $V$ = 0.5 V) and (c) lifted situation ($I$ = 50 nA, $V$ = 0.5 V).
Lines being at identical positions in b and c, mark directions of cross sections along C-C bond directions as shown in (a).
\textbf{d}, \textbf{e} DFT calculated STM image for flat graphene ($V$ = 1 V, constant-height: $z^*$=2 \AA) (d), and buckled graphene ($\Delta z$ = 5.3 pm) at a compression of 16\% (e).
As expected, the buckling leads to an STM image that enhances the sublattice that is shifted towards the STM-tip and reduces the intensity of the other sublattice. Graphene lattice is indicated (yellow lines and dots).
\textbf{f}, Apparent lattice constant in STM images as a function of the local radius of curvature of the long range morphology.
Red curves are a fit of the data points to the intercept theorem, with the effective length of the p$_z$-orbitals (distance of tip from the centre of the C atom) as a free parameter turning out to be $z^* =  5.3$ \AA.
The red dot corresponds to the lifted graphene exhibiting SSB in (c). The green dot shows the required compression for buckling. \textbf{g}, schematic representation of the apparent lattice constant (red double arrow) due to the curvature of the sample.
Blue dots mark the atom core positions, p$_z$-orbitals are visualized by orange clubs. Radius of curvature $R$ is indicated.
\textbf{h}, Buckling amplitude  $\Delta z$ as a function of the in-plane compression as calculated by DFT.
\textbf{i}, Energy per atom as a function of an isotropic compression strain, calculated in LAMMPS \cite{Plimpton1995} with the AIREBO potential \cite{Stuart2000} as a model for the carbon-carbon interaction.}
\label{fig:s6}
\end{figure}

Next we consider possible buckling of graphene, which moves sublattice A upwards and sublattice B downwards geometrically. 
A compression of the atomic lattice could in principle favour such a transition from sp$^2$-bonds to sp$^3$-bonds or, alternatively, to a stable mixture of both bond types.
Then, sublattice A (B) would be closer (further away) from the tip with its p$_z$-orbital pointing towards (away) from the tip orbitals. This leads to the preferential observation of sublattice A.
The required compression might be induced by the flattening of a curved surface during the transition from a valley to a hill, while lifting the graphene (Fig. \ref{fig:s6}a).
This compression is calculated straightforwardly from the geometry to be 0.1\% in Fig. \ref{fig:s6}a and of similar size in all other lifted areas.
The induced strain of  $\sim$0.1 \% interestingly matches the overall strain of the sample found by Raman spectroscopy to be compressive and $\sim$0.1 \% \cite{Lee2012a}.

We have performed ab-initio calculations on the level of density-functional theory (DFT) in the generalized gradient approximation (GGA) in order to check if buckling could explain the observed triangular STM-picture.
The calculations are done with the code Quantum-Espresso \cite{Giannozzi2009}.
The wave-functions are expanded in plane-waves with an energy cutoff at 37 Ry. We have used the projector augmented plane-wave (PAW) method \cite{PhysRevB.50.17953, PhysRevB.59.1758} to describe the core-valence interaction. In this approximation, the equilibrium lattice constant is 2.466 \AA\ (corresponding to a bond-length of 1.424 \AA, slightly overestimating the experimental lattice constant as is usually the case in the GGA).
We compressed the lattice by various amounts and relaxed the geometry in order to check if buckling occurs.
The result is shown in Fig. \ref{fig:s6}h. Up to an (isotropic) compression by 16\%, the planar geometry remains stable
At larger compression (up to 20\%), the sublattice height difference increases quickly and at even larger compression, the buckled planar structure becomes unstable.
We checked by calculations that the presence of a perpendicular electric field (of the order of $\pm$1V/\AA\ - such as it occurs during STM measurements) does not change the threshold for buckling formation.
In principle, buckling induced by compression could thus explain the observed trigonal STM features (see simulated STM-images Fig. \ref{fig:s6}d-e).
However, the applied strain in the measurements is much too small to induce a buckling transition. 

Nevertheless, a priori, one cannot exclude a phase separation into compressed and extended areas within the lifted graphene, which compensate each other in strain. Thus, we estimated the strain observed in the STM images.
This is complicated by the curvature of the graphene topography leading to an apparently larger (smaller) lattice constant on graphene hills (valleys) in STM images \cite{Mashoff2010}.
The reason is the dominating contribution of the p$_z$-orbitals to the tunnelling current.
On curved graphene, the p$_z$-orbitals are tilted with respect to its neighbours, such that, at the tip, neighbouring p$_z$-orbitals are further apart (closer to each other) than at the C atom cores in case of hills (valleys).
The apparent lattice constant measured by STM, i.e. probed at the position of the last tip atom \cite{Tersoff1983}, will therefore be modified by the curvature with respect to the real lattice constant.
We find that this effect can be surprisingly well described by the intercept theorem as sketched in Fig. \ref{fig:s6}g.

Probing the lattice constant in areas, which are barely lifted, as a function of local curvature of the graphene (Fig. \ref{fig:s6}f) fits to the intercept theorem (red lines) within a few percent.
While the determined strain in a lifted area is within these error bars (large red dot in Fig. \ref{fig:s6}f), the required strain of $\varepsilon$ = 16\% (large green dot) is clearly out of the error bar with respect to the upper limit of measured compression of 3\%.
Consequently, a strain of $\varepsilon$ = 16\% can be safely excluded. 

Generally, it might also be possible that a compression pattern is scanned with the tip in a way that does not allow measuring the decreased lattice constant directly.
But then, the scanning tip itself must dominantly induce the compression.
However, it is difficult to imagine that the attractive forces of the tip induce a compressive strain. 
The opposite is the case as shown by our MD.
Moreover, using MD where the atomic interaction in graphene is modelled with the AIREBO potential \cite{Stuart2000}, we find that the in-plane compression of 16\% requires a strain energies $>$2 eV per atom (Fig. \ref{fig:s6}i) to be compared with $\sim$400 mV of tip induced energies to the closest C atom at an unrealistically small tip-graphene distance $z^*$ = 3 \AA.
Thus, the tip forces are not only of the wrong sign, but also too weak to induce a compressive buckling.

\refstepcounter{alter}
\subsection*{S\arabic{supplement}-\arabic{alter}: Strain induced Peierls transition, Kekul\'{e} distortion}
\addcontentsline{toc}{subsection}{S\arabic{supplement}-\arabic{alter}: Strain induced Peierls transition}

\begin{figure}[h!]
\includegraphics[width = 1 \textwidth]{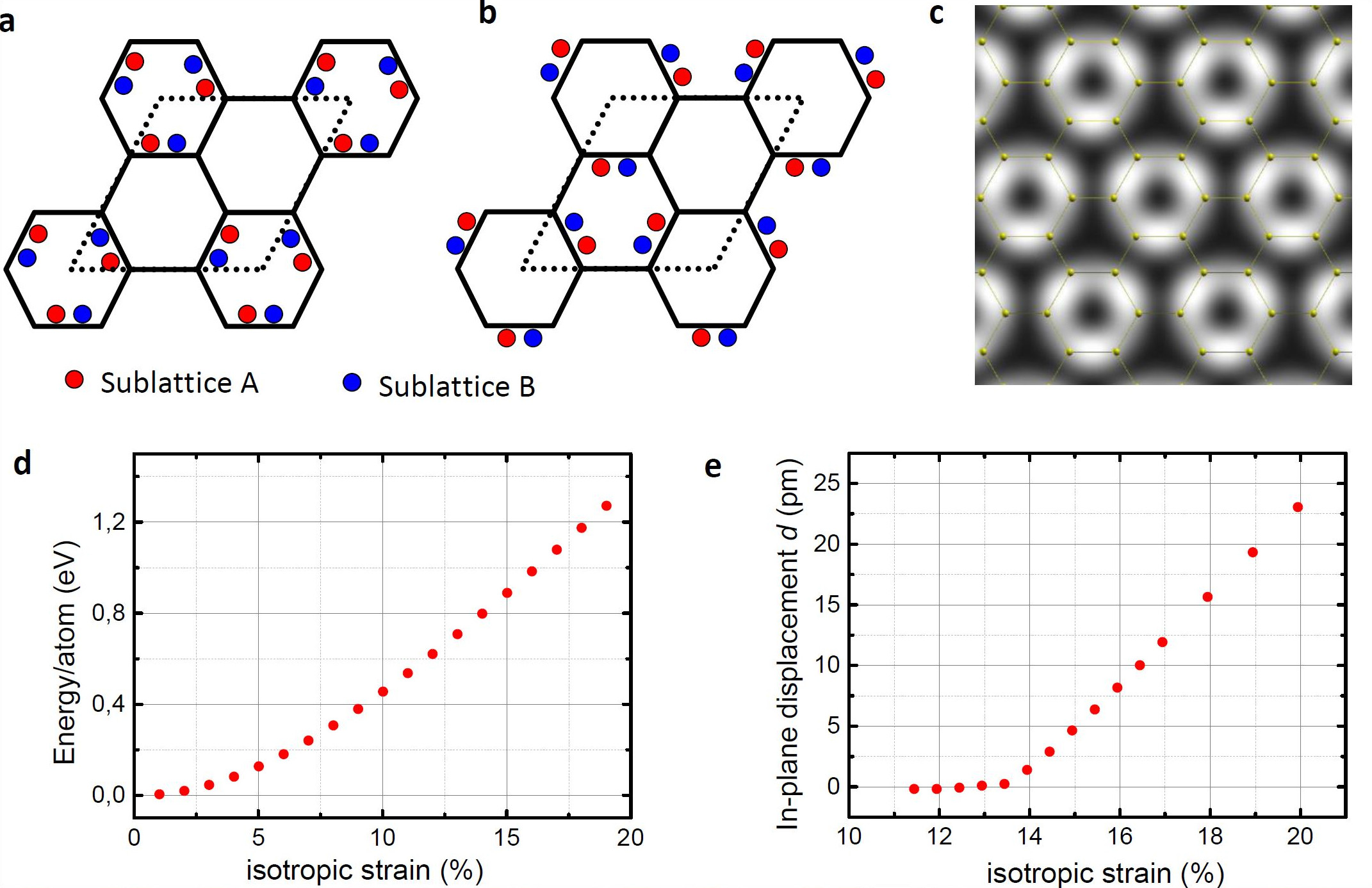}
\caption{\textbf{Strain induced Peierls transition.} \textbf{a}, \textbf{b}, Possible phonon modes in highly (12\%) strained graphene, A1 (a) and B1 (b) with undistorted graphene lattice (black) and unit cell of the distorted calculation cell (dotted black lines) marked \cite{Marianetti2010}. Blue and red dots represent atomic positions at extrema. \textbf{c}, DFT calculated STM image for a relative lateral displacement amplitude of 6 pm between the graphene atoms at the 15.5\% strained graphene lattice in the Peierls phase (A1-mode). The graphene lattice is indicated (yellow lines and dots). \textbf{d}, Energy per atom as a function of an isotropic tensile strain calculated in LAMMPS with the AIREBO potential as a model for the carbon-carbon interaction. \textbf{e}, Calculated lattice deformation $d$ (atomic displacement from the equilibrium position) in A1-mode as a function of isotropic tensile lattice strain.}
\label{fig:s7}
\end{figure}

Another electronic effect that can modify the charge density on the graphene lattice is a Peierls transition, predicted to occur in a real magnetic field, in the quantum Hall regime \cite{Fuchs2007a}.
Since in our experiments we are dealing with a pseudo-magnetic field and the sample does not show Landau levels, we will investigate the effect of a strain induced Peierls transition.
A periodic change of the bond length can make it energetically favourable to adjust the electron system into a charge density wave leading to a gap at the Fermi level in the electronic system and to a softening of the corresponding phonon mode \cite{Marianetti2010}.
A precursor of such a phonon softening is observed in graphene known as the Kohn anomaly at the K point \cite{Piscanec2004, Dubay2003}. However, graphene and graphite do not exhibit a Peierls transition down to lowest temperatures.
With DFT calculations (in agreement with the results of Ref. \cite{Marianetti2010}), we find that biaxial tensile strain can drive the system into a Peierls transition.
However, this requires a large lattice expansion of at least 12\% (Fig. \ref{fig:s7}e). Therefore, similarly to the buckling transition, this is neither compatible with the observed strain nor energetically possible.

Additionally we find, as expected, that the strongest LDOS of the Peierls phase is located between the sublattices, i.e. at the bond sites similar to the Kekul\'{e} phase (Fig. \ref{fig:s7}c).
In contrast, the STM experiment exhibits the largest LDOS (highest positions in constant current mode) at the atomic sites, which can be unambiguously determined, if one observes continuously how the honeycomb appearance at low $I$ transfers into the SSB phase at larger $I$ (Fig. \ref{fig:s6}a-c, blue and red lines).
Thus, we can exclude the Peierls transition as the origin of our SSB.
Furthermore, because the SSB appears at the atomic sites, we can rule out other types of Kekul\'{e} distortions, e.g. brought about by hybridization with the substrate \cite{Gutierrez2016}.

\refstepcounter{alter}
\subsection*{S\arabic{supplement}-\arabic{alter}: $<V, B_{\rm ps}>$ correlation gap}
\addcontentsline{toc}{subsection}{S\arabic{supplement}-\arabic{alter}: $<V, B_{\rm ps}>$ correlation gap}

A final possible reason for the appearance of sublattice symmetry breaking is a correlation of pseudo-magnetic field $B_{\rm ps}$ and a scalar potential $V(r)$, which can be induced by the electric field of the tip \cite{Low2011}. In principle, $V(r)$ can also be induced by strain, but then the required correlation with the strain induced $B_{\rm ps}(r)$ disappears \cite{Low2011}. The Dirac Hamiltonian exhibits a finite mass term ($\propto \sigma_z$), if  $V(r)$ is correlated with $B_{\rm ps}(r)$, i.e. the mass term is roughly proportional to $<V(r), B_{\rm ps}(r)>_r$. This leads to a real gap $\Delta E_{\rm corr}$ and accordingly induces a SSB around  $\Delta E_{\rm corr}$, which continuously weakens at higher energy. Experimentally, we do not observe a gap in $dI/dV$-curves down to, at least, 10 meV, but we find a SSB with nearly voltage independent contrast $C_{\rm exp}$ up to  $V \approx$ 1 V, if the tip-graphene distance is kept constant by adjusting $I$.
This makes this scenario unlikely.

However, since we increase the scalar potential $V(r)$ within graphene with increasing applied bias voltage  $V$, we cannot exclude a priori that $\Delta E_{\rm corr}$ is always smaller than $V$. Notice that one expects a contact potential difference between graphene and tip of about 100 meV \cite{Mashoff2010}, such that a remaining scalar potential is also expected at $V$ = 0 mV implying the persistence of $\Delta E_{\rm corr}$ at low $V$, which was never observed. 

Assuming the unlikely, best case scenario that the tip electric field is perfectly correlated with the pseudo-magnetic field $B_{\rm ps}(r)$, we can use the formula given by Low et al. \cite{Low2011} to estimate an upper bound for the gap $\Delta E_{\rm corr} = B_{\rm ps} V_{\rm el} l^2 e^2 / \hbar$.
Here, $B_{\rm ps}$ and $V_{\rm el}$ is the spatially averaged magnitude of the pseudo-magnetic field and electrostatic potential and $l \approx$ 1 nm is the spatial scale over which the two are correlated.
Plugging in a typical $B_{\rm ps}$ = 1-10 T of rippled graphene on SiO$_2$ \cite{Gibertini2012, Morozov2006} and a scalar potential $V_{\rm el} (r)$ = 0.2 V \cite{Mashoff2010}, we get a gap of  0.3-3 meV at a tip voltage of $V$ = 1 V, indeed much too small to be observed.
However, in the experiment we image SSB up to 35 \% at an energy of 1000 times of such a gap, where any remaining SSB by the correlation effect is negligible ($< 0.1$ \%).
Thus, we safely exclude this scenario as well.

Finally, one can ask if the correlation of $V(r)$ with the induced $B_{\rm ps}(r)$ of the Gaussian deformation is the origin of the SSB.
This cannot be excluded completely. But the fact that $V(r)$ induced by the tip bias $V$ will be at first order rotationally symmetric, while $B_{\rm ps}(r)$ within the Gaussian is sixfold rotationally antisymmetric (Fig. 1c of main text) suppresses the correlation gap significantly.
For a perfect correlation, we find $E_{\rm corr} \simeq 0.6$ eV at $V$ = 1 V. Suppression by a factor of 5 by imperfect correlation would safely exclude this scenario, too, and is geometrically likely.
\\

Closing this chapter, we state that all reasonable explanations for the SSB, with the exception the pseudo-Zeeman effect, strongly fail. 
Together with the quantitative agreement of the strength of the pseudospin-polarization in the effective model with the SSB in the experiment, this provides substantial evidence for the correctness of the pseudospin scenario.

% -------------------------------------------------------------------

\refstepcounter{supplement}
\section*{S\arabic{supplement}: Evaluation of the sublattice contrast \textbf{$\Delta z(x,y)$} and the lifting height \textbf{$H_{\rm exp}(x,y)$}}
\addcontentsline{toc}{section}{S\arabic{supplement}: Evaluation of the sublattice contrast \textbf{$\Delta z(x,y)$} and the lifting height \textbf{$H_{\rm exp}(x,y)$}}
\label{sec:sublcontr}

The experimental LDOS contrast $C_{\rm exp}$ is derived from the measured difference in apparent height between the two graphene sublattices  $\Delta z$ within constant current images. It is evaluated for different tunnelling distances $z$, respectively different tunnelling currents $I$. 

Generally, the determination of $\Delta z$ is disturbed by the corrugation of the long-range morphology, due to the possible finite slope along the A-B bond direction.
It is thus, necessary to remove the long-range morphology (rippling), from the atomic corrugation pattern prior to $\Delta z$ evaluation.

\begin{figure}[h]
\includegraphics[width = 0.8 \textwidth]{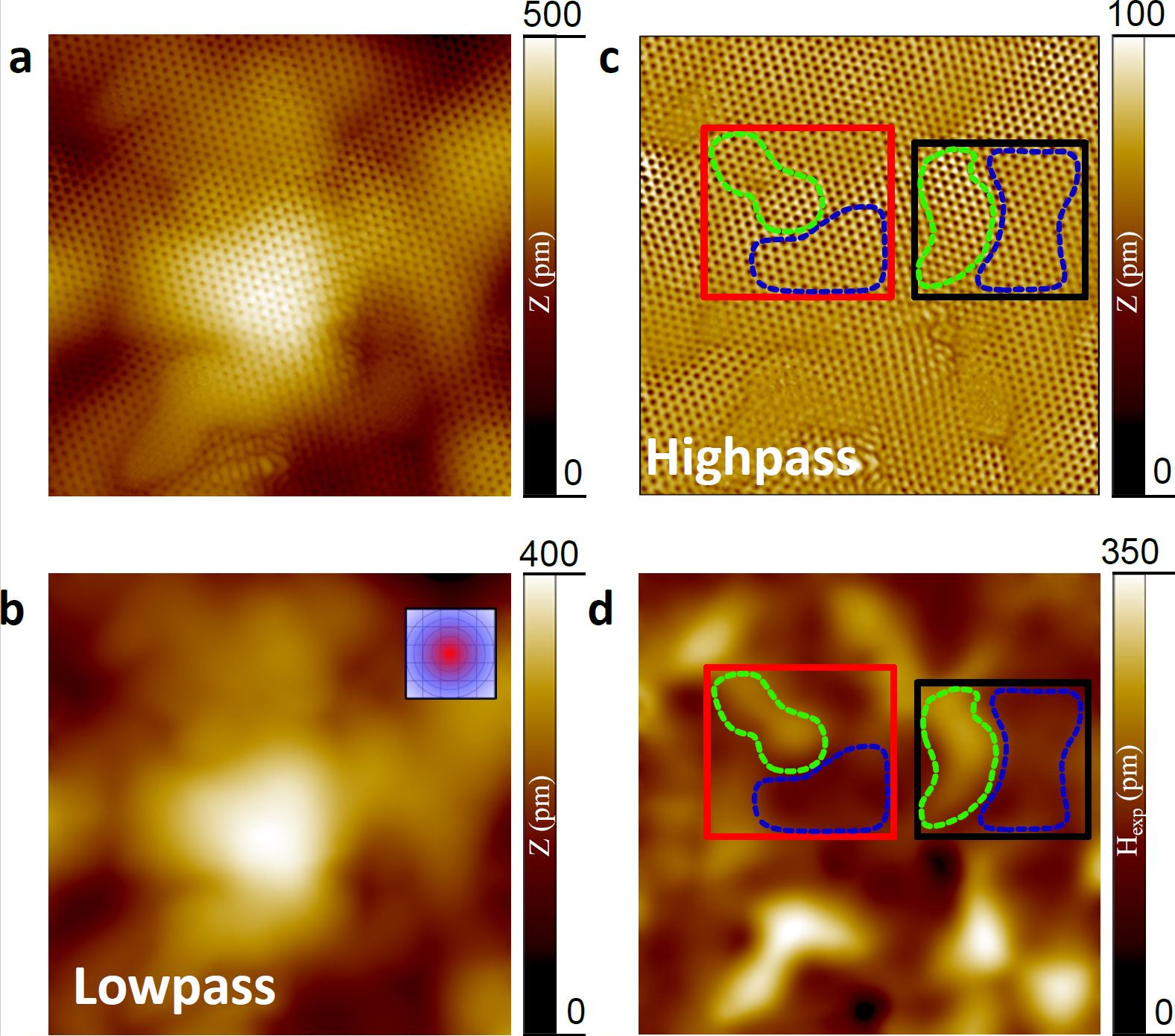}
\caption{\textbf{Determination of $\Delta z$ and $H_{\rm exp}$ from constant current images.} \textbf{a}, Constant current image of graphene on SiO$_2$ (10$\times$10 nm$^2$, $I$ = 50 nA, $V$ = 0.5 V). \textbf{b}, Long-range morphology of (a) deduced by applying a Gaussian smoothing filter as displayed in colour code in the inset. The Gaussian smoothing acts as an effective low pass filter. \textbf{c}, Map obtained by subtraction of (b) from the STM image (a) yielding only the atomic lattice contrast due to an effective high pass filtering. \textbf{d}, Lifting amplitude $H_{\rm exp}$ of graphene at  $I_2$ = 50 nA,  $V$ = 0.5 V derived by subtracting the long-range morphology of the same area recorded at $I_1$ = 0.1 nA,  $V$ = 1 V (barely lifted) from (b) and additionally subtracting homogeneously the required change of tip-graphene distance $\kappa^{-1} \ln(I_2/I_1)$ in order to increase the current, same marks as in (c). Red and black squares mark the areas where $\langle H_{\rm exp} \rangle$ and $\langle \Delta z \rangle$ are evaluated in Fig. 2h of the main text.}
\label{fig:s2}
\end{figure}

Therefore, we firstly apply a Gaussian-weight averaging with large enough Gaussian width in order to remove the atomic corrugation completely.
This leads to the effectively low-pass filtered image in \ref{fig:s2}b exhibiting the rippling only.
Subtracting this from the original image (shown in \ref{fig:s2}a) results in \ref{fig:s2}c exhibiting the atomic corrugation only (See also Fig. 2e of the main text).
Of course, the Gaussian width has to be adapted carefully.
This is done by hand until the atomic corrugation disappears from the low-pass filtered image.
Additionally, $z$-noise on length scales smaller than the atomic corrugation, which is mostly induced by the feedback loop reaction to the lifting of graphene (\ref{fig:s1}), is removed by an additional short-scale Gaussian filter.
The width of this Gaussian is adapted until no atomic corrugations are visible in the removed part of the image, i.e., the full width at half maximum (FWHM) of 75 pm is significantly smaller than the unit cell of graphene.
This procedure is applied to the raw data, e.g., leading to Fig. 1d-g and Fig. 2e of the main text.

After this procedure, $\Delta z$ is determined by profile lines along the C-C bond direction through the image exhibiting the atomic lattice.
In order to determine $\langle \Delta z \rangle$, $\Delta z$ is measured separately for all atom pairs in all three bond directions for areas of relatively constant $H_{\rm exp}$ (marked by green or blue dashed lines in \ref{fig:s2}c and d).
The contrast values $\langle \Delta z \rangle$ shown in Fig. 2h of the main text originate from the areas marked in \ref{fig:s2}c and d.

The local lifting amplitude $H_{\rm exp}(x,y)$ is determined from the height difference between two low pass filtered images (\ref{fig:s2}b) of the same area recorded at high current $I_2$ and low current $I_1$, respectively.
Additionally, $\Delta z = \kappa^{-1}\cdot \ln{(I_2/I_1)}$ is subtracted in order to compensate for the required tip approach towards graphene which increases the current from $I_1$ to $I_2$.
Thereby, $\kappa$ is defined in Eq. 4 of the main text.
Using this procedure, we assume that the image at the lower $I_1$ (0.1 nA) is barely lifted.
\ref{fig:s2}d shows the resulting $H_{\rm exp}$ for $I$ = 50 nA and $V$ = 0.5 V.

In order to display $\langle \Delta z \rangle$ with respect to $\langle H_{\rm exp} \rangle$ (Fig. 2h, main text), $\Delta z$ is measured in selected areas at different $\langle H_{\rm exp} \rangle$, i.e. at different $I$.
The relatively large error bars of $\langle H_{\rm exp} \rangle$ and $\langle \Delta z \rangle$ (Fig. 2h and 3d, main text), are due to the variation of $H_{\rm exp}$ and $\Delta z$ across a selected area, i.e., all other errors of  $\langle H_{\rm exp} \rangle$ and $\langle \Delta z \rangle$ as, e.g., the ones induced by $z$-noise (\ref{fig:s1}e,f), are smaller.

\refstepcounter{alter}
\subsection*{S\arabic{supplement}-\arabic{alter}: Translation of \textbf{$\Delta z$} into the LDOS contrast \textbf{$C_{\rm exp}$}}
\addcontentsline{toc}{subsection}{S\arabic{supplement}-\arabic{alter}: Translation of \textbf{$\Delta z$} into the LDOS contrast \textbf{$C_{exp}$}}
\label{sec:cexp}

Using the averaged sublattice height difference  $\langle \Delta z \rangle$ from a certain area, we deduce the corresponding LDOS contrast $C_{\rm exp}$ as described by Eq. 4 of the main text. Within the Tersoff-Hamann model \cite{Tersoff1983}, the STM current $I$ reads:
\begin{equation}
I \approx \frac{4 \pi e}{\hbar} \int_0^{eV} \nu_{\rm G}(E_{\rm F} - eV + \varepsilon) \cdot \nu_{\rm T}(E_{\rm F} + \varepsilon) \cdot e^{-\kappa z} d \varepsilon
\label{eq:17}
\end{equation}
with  $\nu_{\rm G}$ and  $\nu_{\rm T}$ being the LDOS of graphene and the tip, respectively, and the Fermi energy $E_{\rm F}$ of graphene. For constant tip-graphene distance $z$, an energy-independent change of $\nu_{\rm G}$ on the two sublattices by $+\Delta \nu_{\rm A}$ and  $-\Delta \nu_{\rm B}$, respectively, implies a change of the tunnelling current $I$. The difference of $I$ to the set-point  $I_{\rm S}$ is compensated by a respective adjustment of the tunnelling distance by $\Delta z_{\rm A}$ and $-\Delta z_{\rm B}$, respectively, with $\Delta z = \Delta z_{\rm A} + \Delta z_{\rm B}$. Hence, we find:
\begin{equation}
I_{\rm S,A/B} \approx \frac{4 \pi e}{\hbar} \int_0^{eV} (\nu_{\rm G} \pm \Delta \nu_{\rm A/B})\cdot \nu_{\rm T}\cdot e^{-\kappa (z \pm \Delta z_{\rm A/B})} d \varepsilon
\label{eq:18}
\end{equation}
\begin{equation}
\Rightarrow (\nu_{\rm G} + \Delta \nu_{\rm A})\cdot  e^{-\kappa (z + \Delta z_{\rm A})} = (\nu_{\rm G} - \Delta \nu_{\rm B})\cdot e^{-\kappa (z - \Delta z_{\rm B})}
\label{eq:19}
\end{equation}
In the last step, we reasonably ignore the possible energy dependence of $\kappa$ and $\nu_{\rm G}$.
The LDOS contrast $C_{\rm exp} = \frac{\Delta \nu_{\rm A} + \Delta \nu_{\rm B}}{\nu_{\rm G}}$ is calculated straightforwardly by using  $\Delta \nu_{\rm A} = \Delta \nu_{\rm B}$ as implied by the first order perturbation theory \cite{Tersoff1983}:
\begin{equation}
C_{\rm exp} = \frac{2 \Delta \nu_{\rm A}}{\nu_{\rm G}} = 2 \frac{e^{\kappa \Delta z} - 1}{e^{\kappa \Delta z} + 1}
\label{eq:20}
\end{equation}

\refstepcounter{alter}
\subsection*{S\arabic{supplement}-\arabic{alter}: The effect of the feedback loop in relation to $I(Z)$ curves}
\addcontentsline{toc}{subsection}{S\arabic{supplement}-\arabic{alter}: The effect of the feedback loop in relation to $I(Z)$ curves}
\label{sec:noise}

\begin{figure}
\includegraphics[width = 0.9 \textwidth]{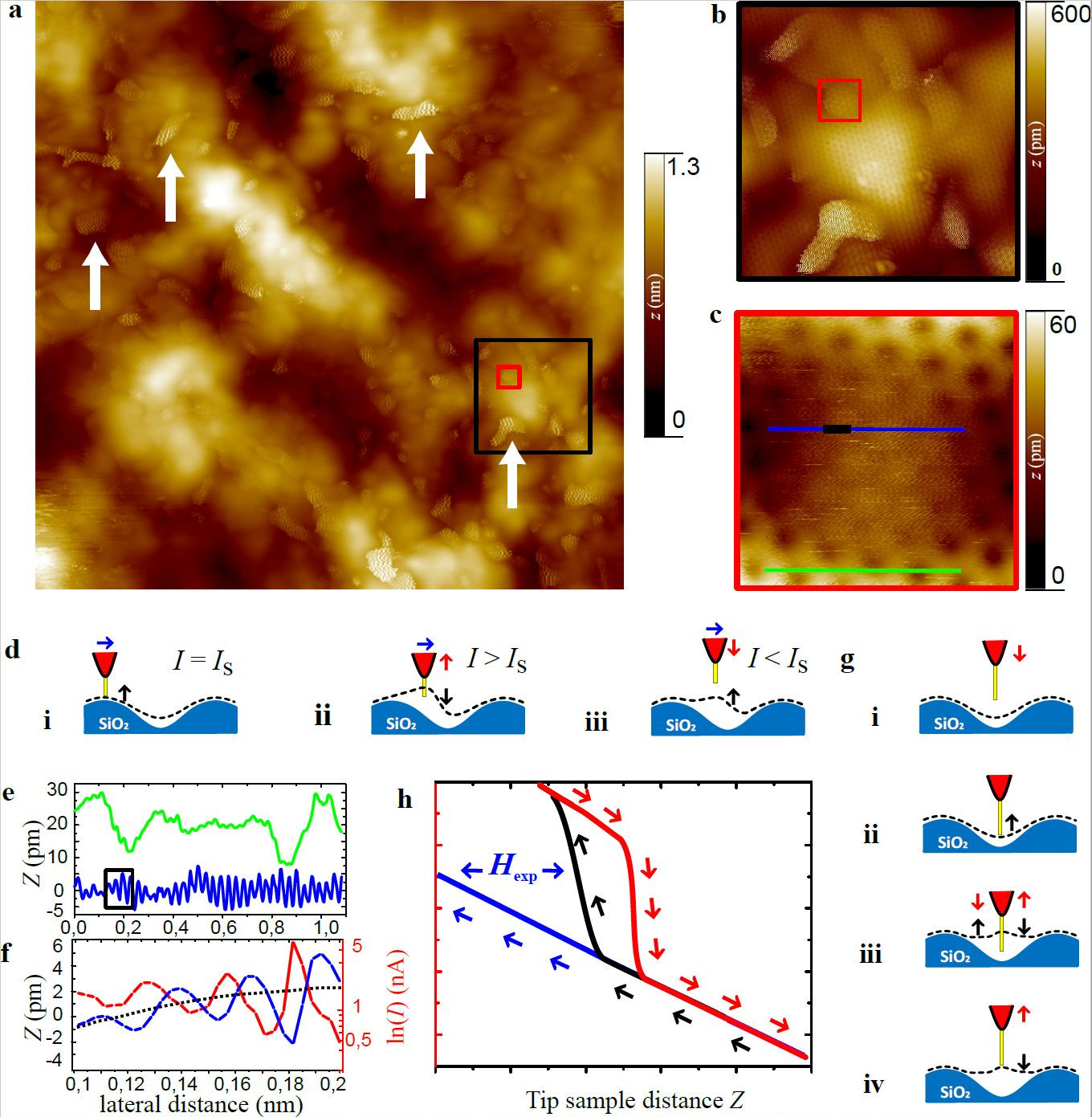}
\caption{\textbf{Interaction between STM feedback and graphene.} \textbf{a}, Constant current STM image of graphene on SiO$_2$, (50$\times$50 nm$^2$, $I$ = 1 nA,  $V$ = -0.3 V), black (red) square marks the zoom area of (b) ((c)); white arrows point to areas of increased $Z$-noise. \textbf{b}, Zoom into (a), (10$\times$10 nm$^2$,  $I$ = 1 nA, $V$ = 0.5 V), red square marks area of (c). \textbf{c}, Zoom into (b) (1.5$\times$1.5 nm$^2$, $I$ = 1 nA, $V$ = 0.5 V), blue and green (black) stripes mark profile lines displayed in (e), ((f)). \textbf{d}, Sketch of the tip (red) above partially suspended graphene (dashed line) deposited on SiO$_2$ (blue). 
Arrows mark the lateral tip movement (blue), the tip induced movement of graphene (black), and the vertical movement of the tip induced by the feedback correction (red). The yellow bar indicates the required tip-graphene distance for  $I = I_{\mathrm{S}}$: (\textbf{i}) tip on supported area. (\textbf{ii}) initial tip position on suspended area, (\textbf{iii}) tip position on suspended area after feedback induced tip retraction. \textbf{e}, Profile lines from (c) through a lifted area (blue) and a supported area (green). Black square marks the zoom shown in (f). \textbf{f}, Feedback induced  $Z$-movement (blue) and simultaneously measured logarithmic tunnelling current (red) from the zoom of (e); the $Z$-movement follows $\ln{(I)}$ with a time delay; note that the displayed lateral distance is less than a graphene lattice constant such that only part of an individual atom of one sublattice is probed as profiled by the dotted line.
\textbf{g}, Graphene movement during  $I(Z)$-curve, same meaning of colours and symbols as in (d): (\textbf{i}) tip approach, (\textbf{ii}) begin of lifting, (\textbf{iii}) reaction of graphene to tip approach (left arrows) and tip retraction (right arrows), (\textbf{iv}) possible hysteresis of graphene lifting after tip retraction. \textbf{h}, Schematic  $I(Z)$-curves with hysteresis during tip approach (black) and retraction (red). A comparison with the  $I(Z)$-curve expected on a vertically fixed substrate (blue) is used to measure the lifting amplitude $H_{\rm exp}(Z)$ of graphene as marked.}
\label{fig:s1}
\end{figure}

\ref{fig:s1}a shows a constant current STM image of graphene on SiO$_2$ with zooms displayed in \ref{fig:s1}b and c.
Several areas, some of them marked by arrows, exhibit an enhanced noise in the topography.
The enhanced noise indicates that these areas are more strongly lifted by the forces of the STM tip as cross-checked by $I(Z)$-curves.
The $Z$-noise is due to the fast vertical retraction of the STM tip by the feedback loop, which is triggered by the suddenly increased current during the lifting of graphene (\ref{fig:s1}d).
We find that the amount of noise depends on the feedback parameters, such as the bandwidth with respect to the recording time per pixel and the stabilization current.
The noise also spatially varies at given feedback parameters, which we ascribe to different local lifting amplitudes  $H_{\rm exp}$ induced by a different strength of the local adhesion between graphene and the substrate.
Since the bandwidth of our feedback loop (1 kHz) is much lower than the eigenfrequency of the graphene membrane ($\sim$1 THz) \cite{Mashoff2010}, we get a retarded reaction of the STM-servo, such that the $Z$-correction overshoots, leading to an enhanced $Z$-noise (\ref{fig:s1}d-f).
Since a larger  $H_{\rm exp}$ leads to a stronger current $I$ and thus, to a stronger retraction of the tip by the feedback, a large  $H_{\rm exp}$ implies a large $Z$-noise.
In turn, the $Z$-noise is a fingerprint of the local adhesion force between graphene and the substrate.

During  $I(Z)$-curves, the tip is firstly approached towards graphene and retracted afterwards, while the feedback loop is switched off.
The resulting movement of tip and graphene is sketched in \ref{fig:s1}g including a possible hysteresis of the graphene lifting \cite{Mashoff2010}.
\ref{fig:s1}h sketches the resulting $I(Z)$-curve with hysteresis. Such a hysteresis is partially also found in the experimental $I(Z)$-curves \cite{Mashoff2010} and within the MD (not shown).
Measuring $I(Z)$ with feedback loop on, i.e. deducing $Z$ from a series of constant current images at different $I$, corresponds to a situation between an approaching and a retracting  $I(Z)$-curve without feedback loop.
Consequently, the $I(Z)$ values probed with feedback loop are larger than the $I(Z)$ values recorded without feedback loop during the approach.
This is indeed found as visible in Fig. 2a of the main text.

\refstepcounter{alter}
\subsection*{S\arabic{supplement}-\arabic{alter}: Sublattice symmetry breaking over large areas}
\addcontentsline{toc}{subsection}{S\arabic{supplement}-\arabic{alter}: Sublattice symmetry breaking over large areas}

\begin{figure}[h!]
\includegraphics[width = 1 \textwidth]{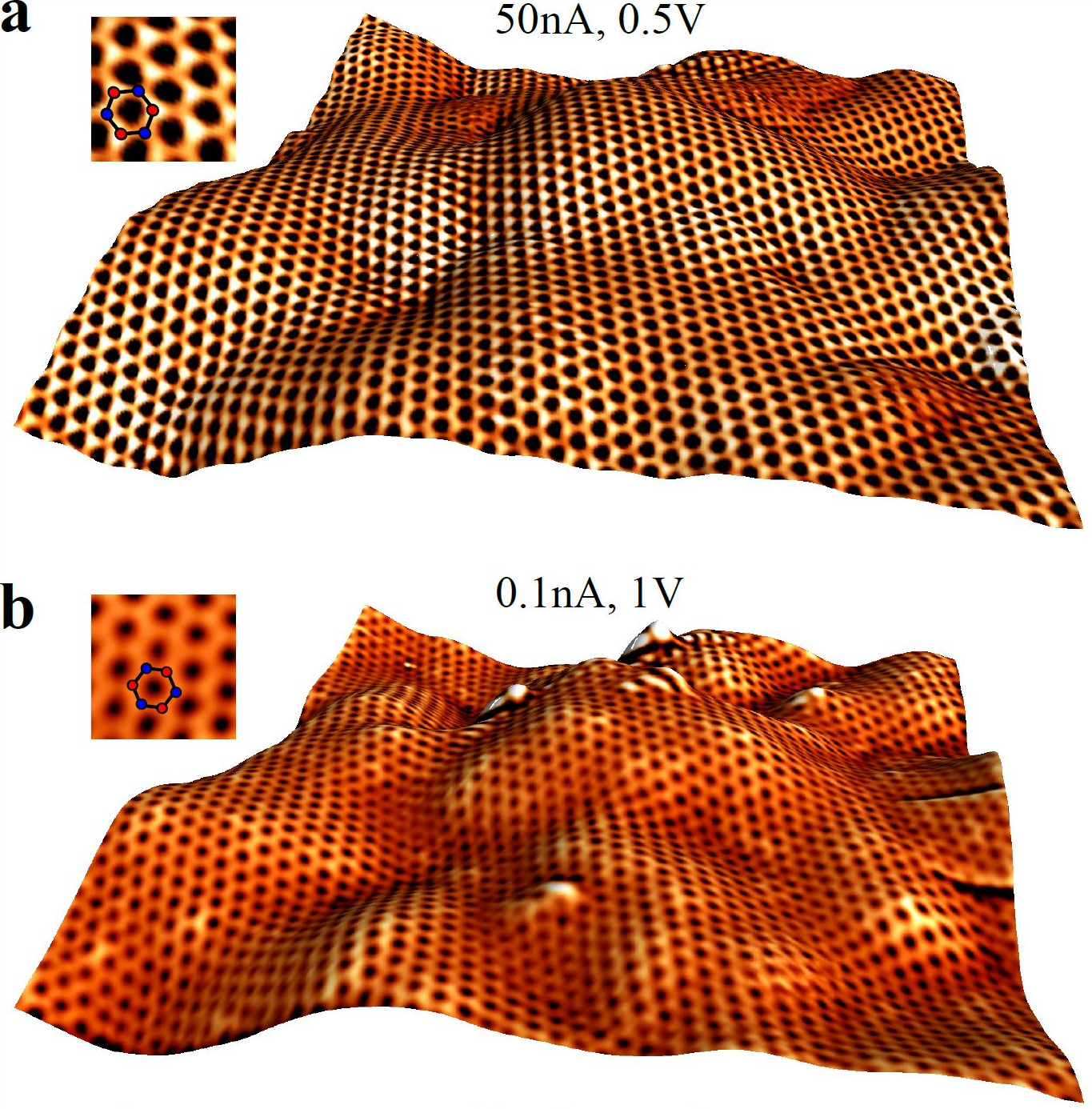}
\caption{\textbf{Sublattice symmetry breaking on a 10$\times$10 nm$^2$ area:} Constant current STM images (10 nm$^2$) of graphene on a SiO$_2$ substrate. The 3D topography results from the long range Gaussian convolution filtering (FWHM = 3 \AA), showing the large scale graphene topography. The colour code results from the subtraction of the topography from the raw data, showing the atomic corrugation.
\textbf{a}, The whole graphene membrane is lifted resulting in a smoothing of the corrugation induced by the SiO$_2$ substrate. The whole area shows SSB, with the sublattice marked red being higher (see inset). The SSB $\langle \Delta z \rangle$ is in the 10 to 20 pm range, depending on lifting amplitude.
\textbf{b}, STM image at low tunnelling current of the same area as in (a), showing the honeycomb atomic lattice of the graphene membrane in most areas. The graphene is expected to be barely lifted (see $I(Z)$ curve at 0.1 nA in Fig. 2a of the main text). The morphology shows a much higher corrugation amplitude compared to (a), resulting from the surface roughness of the SiO$_2$ substrate.}
\label{fig:SSBlarge}
\end{figure}

Our model described in the main text implies that, as long as the STM tip remains unchanged, the tunnelling tip will scan within an area of constant sign of $B_{\rm ps}$ (see supplementary video).
This means that the same sublattice will appear higher, all over the sample.
In Fig. \ref{fig:SSBlarge}a we show a large area (10$\times$10 nm$^2$), measured at large tunnelling current (50 nA).
The atomic resolution image clearly shows one of the sublattices being higher (marked red) all over the sample surface.
The magnitude of the sublattice contrast can change as a function of the local lifting height (see Fig. 2d, e of the main text).
As a comparison, a constant current image of the same area is shown in Fig. \ref{fig:SSBlarge}b, measured at low tunnelling current / low lifting.
It displays the honeycomb lattice of graphene, with the sublattices having equal height in most areas of the image.

\newpage

\newpage

\refstepcounter{supplement}
\section*{S\arabic{supplement}: Tight-binding calculation and contrast evaluation of sublattice symmetry breaking}
\addcontentsline{toc}{section}{S\arabic{supplement}: Tight-binding calculation and contrast evaluation}

The tight binding calculations were carried out, as described elsewhere \cite{Carrillo-Bastos2014}.
The Gaussian deformation was implemented by a position dependent nearest neighbour hopping parameter while total and local density of states (LDOS) were obtained using recursive Green’s function methods.
Due to the ribbon geometry used for the tight binding (TB) calculation, the LDOS is strongly affected by the boundary conditions, showing spatial oscillations at the atomic scale even for undeformed ribbons \cite{Carrillo-Bastos2014}.
These finite size effects produce sublattice symmetry breaking SSB on zigzag terminated ribbons (caused by the boundaries) and streamline currents in the armchair ribbons (AGNR) \cite{Wilhelm2014}.
Hence we consider an AGNR with width of 11 nm and length of 12 nm.
In order to eliminate the finite size effects, Fast Fourier transform methods were used to filter the associated finite momenta contribution, which are similarly observed in ribbons with and without the Gaussian deformation.
The filtered data was Fourier transformed back to real space, where the LDOS at different sublattice sites was determined for ribbons with Gaussian deformations.
This filtering method might influence the absolute values of the SSB, but since it is applied identically to the different deformations, the relative $\Delta z$ values are barely influenced by the procedure. 

To simulate the sublattice contrast observed by STM, we calculate LDOS data from ribbons with different central positions of the Gaussian deformation within the graphene lattice.
Two examples of this calculation can be seen in Fig. 3e and f of the main text.
In these figures the colour scale encodes the LDOS difference which results from subtracting the LDOS of the pristine AGNR from the one containing the Gaussian deformation.
In these images Fourier filtering was not used.
In order to simulate the tip scanning, for each position of the Gaussian within the AGNR, only the LDOS at a constant distance from the centre of the Gaussian towards armchair direction is plotted (Fig. 3g of main text).
All these calculations were repeated for different LDOS energies, values of the elastic parameter $\beta$, system sizes, and deformation sizes. The system size did not change the observed SSB contrast, while it is found to be proportional to  $\beta$ ($\beta$ = 3 in main text) as expected from Eq. 3 of the main text.

%--------------------------------------------------------------------------

\refstepcounter{supplement}
\section*{S\arabic{supplement}: Sublattice symmetry breaking in a graphene bubble}
\label{sec:bubble}
\addcontentsline{toc}{section}{S\arabic{supplement}: Sublattice symmetry breaking in a graphene bubble}

If we sacrifice the tunability of the strain, available through lifting the graphene by the tip, the presence of SSB can be checked in static graphene deformations.
Within the literature there are numerous observations of SSB in strained graphene, measured by STM \cite{Lu2012, Xu2009, Sun2009, Gomes2012}.
One intriguing example is the paper by Lu et al. \cite{Lu2012}, where bubbles of graphene are prepared on a Ru substrate.
They show that regions of the bubbles having low strain show a honeycomb atomic structure, while regions with high strain have a sublattice symmetry broken atomic lattice (Fig. 3c, 4h and S8 in ref. \cite{Lu2012}). 
However, the authors don't explain the origin of the SSB.

Of course, the presence of strain does not necessarily mean that there is a finite $B_{\mathrm{ps}}$ field present.
Therefore, to check for increased SSB in static graphene deformations, we have studied bubbles on a sample of graphene supported on hexagonal boron nitride (BN).
The stacking of graphene onto BN is known to result in the formation of bubbles below the graphene and probably containing hydrocarbons \cite{Haigh2012}.
Usually these bubbles are too large for stable STM imaging, having lateral sizes in the 10 nm to 1 $\mu$m range.
However, by STM measurements on a dry stacked graphene/BN sample \cite{Freitag2016} we have identified a bubble having a width ($b$) of 5.2 \AA\ and 8.5 \AA\ in two perpendicular directions and a height ($H$) of 2.28 \AA.
This is similar to the size of the deformation induced by the STM tip on SiO$_2$.

\begin{figure}[h]
\includegraphics[width = 1 \textwidth]{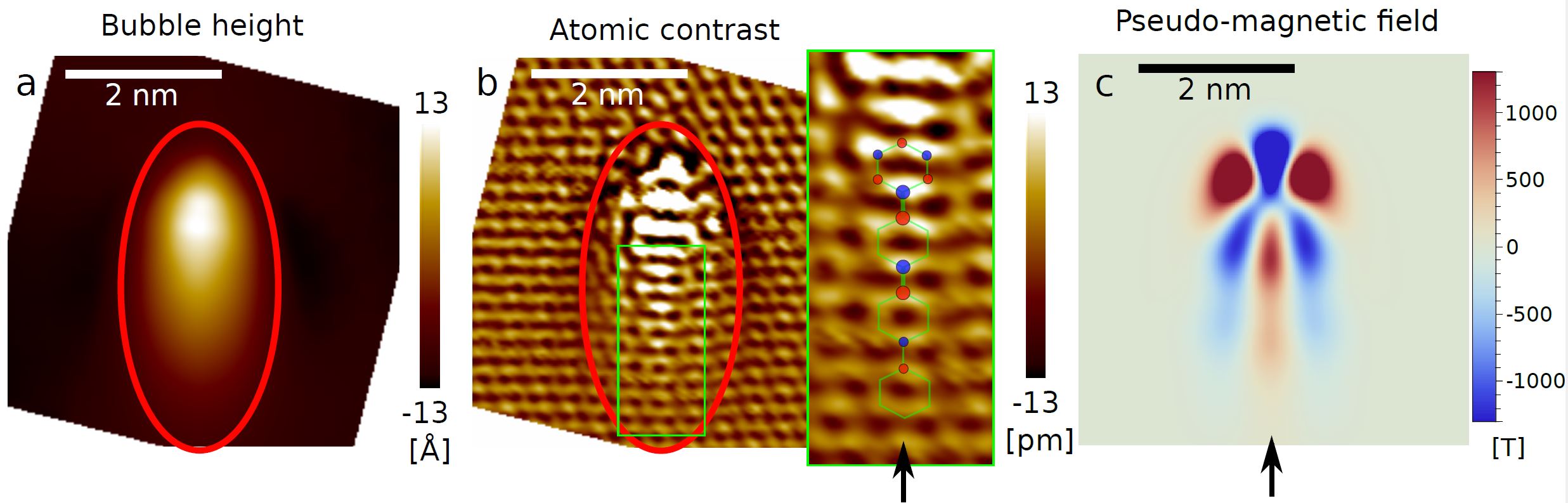}
\caption{\textbf{Graphene bubble on hexagonal boron nitride.}
Red ellipse marks the graphene bubble. The STM imaging parameters are, 0.4 V and 0.4 nA.
\textbf{a}, Long-range morphology of the bubble, obtained by low-pass filtering of the STM image.
\textbf{b}, Atomic corrugation on the graphene bubble, obtained by high-pass filtering. Inset: zoom of the area marked by the green rectangle. Moving from the BN up along the bubble ridge, the sublattice symmetry breaking due to the BN (bottom) is inverted and gradually increases due to the increase of $B_{\rm ps}$ (the sublattice marked blue being higher).
\textbf{c}, Pseudo-magnetic field pattern of the graphene bubble, calculated by fitting four Gaussians to the graphene bubble.
}
\label{fig:bubble}
\end{figure}	

The bubble (Fig. \ref{fig:bubble}) is rotationally not symmetric, with a ridge along the armchair direction (shown by black arrows).
Its orientation with respect to the armchair direction is a favourable coincidence, since it allows for an extended pseudo-magnetic field along the ridge of the bubble of 400-1000 T (Fig. \ref{fig:bubble}c).
Indeed if we examine the SSB in the atomic resolution image (Fig. \ref{fig:bubble}b), we observe an increasing SSB as $B_{\mathrm{ps}}$ increases. 
Starting from the bottom of the ridge (\ref{fig:bubble}b inset) the atoms marked red are measured to be higher, due to the influence of the BN support. This SSB inverts and becomes stronger towards the top of the bubble, with the atoms marked blue being higher.
Measured as the height difference ($\Delta z$) between the blue and red atomic positions, it has values of 8.4 pm and 5.3 pm, at the site of the vertical bonds marked by larger blue and red atoms in the inset of Fig. \ref{fig:bubble}b.
This is similar to the SSB observed in Fig. 2h of the main text for a lifting height of 1.8 \AA.
The resulting LDOS contrast has values of 18\% and 11\%.
Calculating the LDOS contrast in tight-binding for the same two bonds, we end up with values of 15\% and 5\%, being reasonably consistent.
In comparing these numbers, one should consider that the SSB on the bubble could be influenced by the additional strain components in the graphene on BN, as well as by adsorbates trapped between the graphene and BN.

%--------------------------------------------------------------------------

\refstepcounter{supplement}
\section*{S\arabic{supplement}: Valley filter}
\addcontentsline{toc}{section}{S\arabic{supplement}: Valley filter}

\begin{figure}[h]
\includegraphics[width = 1 \textwidth]{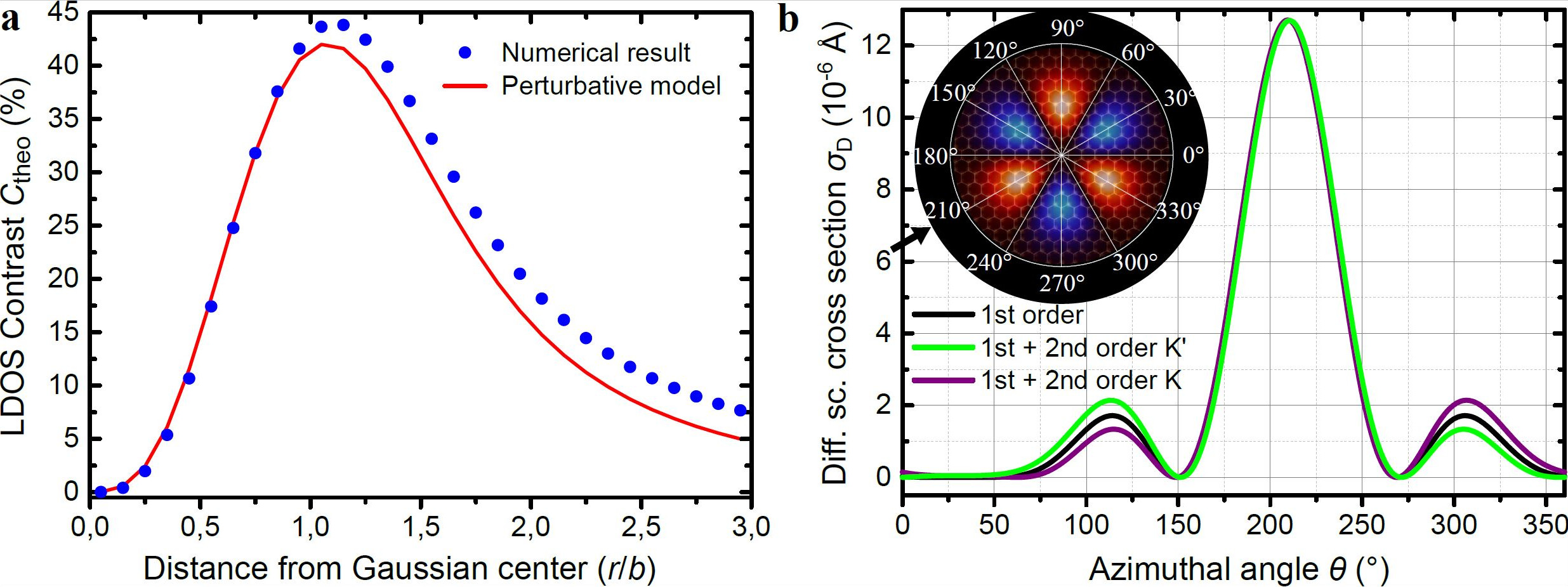}
\caption{\textbf{Cross check of perturbative calculation and valley filter calculation.} \textbf{a}, LDOS contrast between sublattices $C_{\rm theo}$ calculated within the effective Dirac model for a deformation of dimensions $H$ = 2.5 \AA\ and  $b$ = 5 \AA, as a function of radial distance $r$ (measured in units of $b$) for a fixed azimuthal angle of $\theta = \pi /2$. The figure shows the comparison between the perturbative expression (Eq. (3), main text) and the exact result calculated numerically using scattering matrix methods \cite{Schneider2015}. \textbf{b}, Differential scattering cross section for plane-wave spinors originated at valleys K (purple) and K' (green) calculated up to second order perturbation theory in the Lippmann-Schwinger equation formalism. Inset shows pseudo-magnetic field lobes of the K-valley as colour code for a Gaussian deformation with parameters $H$ = 1 \AA\ and  $b$ = 5 \AA\ and the azimuthal scattering angle $\theta$. The incident plane-waves move parallel to the armchair direction (shown by black arrow,  $\theta$ = 210$^\circ$) at energy $E$ = 300 meV. Black curves correspond to first order corrections and show identical contributions from both valleys (non-unitary scattering matrix). Second order corrections reveal that states originating from valleys K (purple) and K' (green) deflect differently by 40 \% with additional differences of 1.5$^\circ$ in electron trajectories.}
\label{fig:s5}
\end{figure}

As stated in the main text, the inhomogeneous nature of the pseudo-magnetic field produced by the deformation might lead to different deflection directions of electrons from K and K' valleys.
Generally, if an electron hits one lobe of the $B_{\rm ps}$ pattern (inset in Fig. \ref{fig:s5}b), it acquires a valley dependent deflection.
This deflection might guide the electrons in other lobes of the same sign of $B_{\rm ps}$, thereby substantiating the deflection. 
Thus, electrons from the K valley might be deflected to the left and electrons from the K' valley to the right.\\

However, the largest fields found in our experiment (Fig. 3d, main text) are up to 4000 T implying magnetic lengths of $l_{\rm B}> 0.4$ nm.
Thus, the cyclotron diameter is always larger than $b$, depending in detail on $H$ and $b$ of the deformation as well as on the LDOS energy of the incoming electron.
This large cyclotron diameter relative to $b$ also explains why we do not observe any orbital quantization (Landau levels) within the deformation.

Nevertheless, by tailoring of $H$, $b$ and electron energy, an effective valley filter, which exploits the valley degree of freedom for information processing, can be constructed.

To investigate the valley filter characteristics, we used a standard elastic scattering formalism based on the Lippmann-Schwinger equation applied to the Dirac equation.
In this approach the strain field is represented by a pseudo-magnetic vector potential \cite{CastroNeto2009} that gives rise to the term  $-e v_{\rm F} \vec{\sigma} \vec{{\cal{A}}}_{\rm ps}$ within the Dirac Hamiltonian treated in perturbation theory.

The differential scattering cross sections for a plane wave pseudospinor (eigenstate of the undeformed Hamiltonian) injected along the armchair direction (see inset in Fig \ref{fig:s5}b), are obtained for $g_{\rm v} (H/b)^2$ = 280 meV $< E$, with  $g_{\rm v} \sim$ 7eV and in the low-energy scattering regime $kb \ll 1$, up to second order in perturbation theory.
To distinguish the contributions from each valley, we chose two pseudospinors with energy $E$ and momentum $\vec{k}$ measured with respect to valleys K and K' with the same velocity (defined as $\nabla_{\vec{k}} E$), i.e., two eigenstates not related by time-reversal invariance, yet moving in the same direction.

The results show that first order (Born approximation) corrections do not distinguish contributions from K- and K'-valleys \cite{Yang2012a}, a fact that can be attributed to the lack of unitarity of the scattering matrix at this order.
However, second order terms do reveal different contributions from each valley that strongly depend on the orientation of motion of the incident state. Fig. \ref{fig:s5}b shows the differential cross section for two plane-wave pseudospinors from K and K' valleys, incident along the armchair direction of the graphene lattice, at $E$= 300 meV.
A strong backscattering ($\simeq 75$ \% at $210^\circ$) is found and opposite preferential deflections by about $100^\circ$ for the two valleys (peaks at $113^\circ$ and $306^\circ$).
These deflections are the consequence of the anisotropic spatial distribution of the pseudo-magnetic field.
Analysis of Fig. 1c in the main text, shows that a pseudospin state originating from the K valley experiences a net pseudofield value of a positive sign, while the one that originating from the K' valley experiences a net field of negative sign. 
The difference results in a net valley polarization of 40\% for parameters of the STM induced deformation on originally supported areas.
Analysis of data for different incident orientations confirms that the valley polarization effect is strongest in the regions with maximal pseudo-magnetic field (as shown).

% \bibliographystyle{nature}
% \bibliography{sup_references_merged-with-main}

\end{document}